\documentclass[%
 reprint,
superscriptaddress,
groupedaddress,
runinaddress,
 amsmath,amssymb,
 aps,
prb,
]{revtex4-1}

\usepackage{algorithm}
\usepackage{algpseudocode}
\usepackage{array}

\usepackage{graphicx}% Include figure files
\usepackage{dcolumn}% Align table columns on decimal point
\usepackage{bm}% bold math
\usepackage{multirow}

\usepackage[table]{xcolor}
\usepackage{textcomp}
\usepackage{booktabs} % For \toprule, \midrule, \bottomrule
\usepackage{tabularx} % For tabularx environment
\usepackage{ragged2e} % For better text justification with tabularx (optional, but good practice)
\usepackage{colortbl} % For \cellcolor
\usepackage{hhline} % Crucial for full cell coloring with tabcolsep

\usepackage{subcaption}

\newcommand{\changed}[1]{ {{ #1 }}}

\newcommand{\Loss}{\mathcal{L}}

\begin{document}
\title{Graph atomic cluster expansion for foundational machine learning interatomic potentials}

\author{Yury Lysogorskiy}\thanks{Corresponding author: yury.lysogorskiy@rub.de}
% \affiliation{Interdisciplinary Centre for Advanced Materials Simulation (ICAMS), Ruhr-University Bochum, 44780  Bochum, Germany}

%\email[Corresponding author: ]{yury.lysogorskiy@rub.de}
\author{Anton Bochkarev}
% \affiliation{Interdisciplinary Centre for Advanced Materials Simulation (ICAMS), Ruhr-University Bochum, 44780  Bochum, Germany}

\author{Ralf Drautz}
\affiliation{Interdisciplinary Centre for Advanced Materials Simulation (ICAMS), Ruhr-University Bochum, 44780  Bochum, Germany}

\begin{abstract}
Foundational machine learning interatomic potentials that can accurately and efficiently model a vast range of materials are critical for accelerating atomistic discovery. We introduce universal potentials based on the graph atomic cluster expansion (GRACE) framework, trained on several of the largest available materials datasets. Through comprehensive benchmarks, we demonstrate that the GRACE models establish a new Pareto front for accuracy versus efficiency among foundational interatomic potentials. We further showcase their exceptional versatility by adapting them to specialized tasks and simpler architectures via fine-tuning and knowledge distillation, achieving high accuracy while preventing catastrophic forgetting. This work establishes GRACE as a robust and adaptable foundation for the next generation of atomistic modeling, enabling high-fidelity simulations across the periodic table.

\end{abstract}

\date{Originally submitted: August 26 2025; Revised: January 3 2026}

\maketitle

\section{Introduction}

The ability to predict materials properties from atomistic simulations is essential for modern materials design. Machine learning interatomic potentials (MLIPs), trained on data from electronic structure methods like density functional theory, have recently emerged as a powerful tool, achieving excellent accuracy for diverse systems \cite{Behler07,Bartok2010,Shapeev16,Drautz19,unkeMachineLearningForce2021,musaelianLearningLocalEquivariant2023,jacobsPracticalGuideMachine2025,kulichenkoDataGenerationMachine2024,WANG2024109673,Thiemann_2025,yuan2025foundationmodelsatomisticsimulation}. 
However, the applicability of conventional MLIPs is constrained by their limited elemental scope. Extending an existing MLIP to include a new chemical element requires generating thousands of new reference calculations and retraining the entire model - a process that represents a significant computational bottleneck.

Foundational interatomic potentials seek to resolve this bottleneck by creating a single, universal model that encompasses the entire periodic table from the outset. This ambitious goal requires both enormous training datasets and an efficient method for representing the vast space of chemical interactions. A brute-force enumeration of interactions is unfeasible; parameterizing just the four-body interactions 
would involve on the order of $10^8$ interactions.

Foundational MLIPs resolve this problem by embedding complex chemistry into a low-dimensional space; a long-standing concept in materials simulation. Early tight-binding models, for example, used the valence electron count to effectively describe chemical trends and structural stability across multiple elements \cite{pettiforTheoryCrystalStructures1970}. Modern  foundational MLIPs build upon this legacy, using multi-dimensional embeddings to leverage the inherent correlations between elements. This approach is remarkably effective, enabling the entire periodic table's chemistry to be captured in few dimensions.\cite{DGPettifor_1986,pettiforbook,SEISER2011749,acs.chemmater.5b04299, qua.24917,Parsaeifard_2021,PhysRevMaterials.7.045802,cerqueira2025nonorthogonalrepresentationchemicalspace}

The first universal force field was published more than 30 years ago\cite{rappeUFFFullPeriodic1992}. Parameterizations of MLIPs across the periodic table started to appear when large training datasets became available \cite{chen2022universal,dengCHGNetPretrainedUniversal2023,batatiaFoundationModelAtomistic2024,barroso2024open,yang2024mattersim,kim2024data,yin2025alphanet,zhang2025graph,fu2025learning,mazitovPETMADLightweightUniversal2025,liang2025nep89}.

To date the development of universal MLIPs was tightly bound to the progress in message passing graph neural networks \cite{Klicpera2020DimeNet,Anderson2019Cormorant,Lubbers2018,Thomas2018TensorField,Batzner2022nequip,Satorras2021EGNN,Unke2019Physnet,Schutt2017schnet,Haghighatlari2021newtonnet,Schutt2021Painn,Batzner2022nequip,Klicpera2022gemnet,Chmiela2017machine,Pozdnyakov2023smooth,Nigam2022unified,Batatia2022design,bochkarev2022multilayer,bochkarev2024}, despite the fact that universal parameterizations across the periodic table are in principle independent of MLIP architecture. The first universal MLIPs that built on the Atomic Cluster Expansion (ACE) \cite{Drautz19} became available two years ago, within the framework of Multi-ACE  \cite{batatiaDesignSpaceE3equivariant2025} that employs general many-body messages and as implemented in MACE \cite{batatia2022mace,batatiaFoundationModelAtomistic2024}.

Here we employ the Graph Atomic Cluster Expansion (GRACE)\cite{bochkarev2024}. By extending ACE to tree-graphs, GRACE stands out from graph neural networks by providing a complete basis  for the parameterization of atomic interactions as a function of atomic positions and chemical species. By straightforward tensor decomposition of the GRACE expansion coefficients one directly obtains sparse representations with efficient chemical embedding that can be evaluated recursively. The recursive evaluation of graph basis functions can be understood as message passing and because of its complete basis, GRACE is able to rationalize and represent other message passing graph neural networks architectures in general. As GRACE further facilitates recursive evaluation of effective ACE on each message passing layer, it benefits from linear scaling with the number of recursion layers as well as linear scaling with the complexity of ACE messages within each layer for efficient double-recursive evaluation.

Training data for universal force fields needs to cover the periodic table comprehensively. Only few publicly available datasets are suitable, notably the Materials Project~\cite{jain2013commentary,deng2023chgnet}, Alexandria~\cite{schmidt2023machine,wang2023symmetry} and the OMat24~\cite{barroso_omat24} datasets that are in the focus of our work here, but also the Open Quantum Mechanical Database~\cite{kirklin2015open} (OQMD) and AFLOWLIB~\cite{curtarolo2012aflowlib} and more recent additions such as MatPES~\cite{MATPES} and MP-ALOE~\cite{kuner2025mpaloe}.

Validation of universal force fields is challenging. Traditional strategies that are employed for MLIPs with only few elements and that probe test errors for specifically relevant simulation tasks are not possible because of the combinatorically many different simulations that would be required across the periodic table. Validation therefore is necessarily limited to tests that seem particularly relevant or are widely adopted in the community. 

In this work we present a number of GRACE models with varying complexity that were parameterized on the OMat24~\cite{barroso_omat24}, Alexandria~\cite{schmidt2023machine,wang2023symmetry} and MPTraj~\cite{deng2023chgnet} datasets and can serve as a foundation models for atomistic modelling in materials science. 

\section{Results}

\subsection{Foundational GRACE interatomic potentials}

Developing foundational MLIPs capable of accurate predictions across a wide array of chemical elements and diverse structures requires exceptionally large and varied datasets. The GRACE framework is designed to effectively manage this inherent complexity.
Our primary training source was the OMat24 dataset~\cite{barroso_omat24}, which currently is the largest publicly available compilation for materials property prediction.
It encompasses 110 million DFT calculations, primarily computed with VASP~\cite{kressabinitio1993,kresse1996efficiency,kresse1996efficient} with the GGA-PBE functional\cite{Perdew1996}, including Hubbard U corrections for specific oxides and fluorides, consistent with Materials Project defaults~\cite{jain2013commentary}. 
Importantly, OMat24 extends beyond near-equilibrium structures, distinguishing it from datasets like Alexandria and MPTraj, but covers same 89 elements as those two. 
The dataset's diversity comes from its generation methods, which include Boltzmann sampling of structures with randomly displaced atomic positions, ab initio molecular dynamics (AIMD), and subsequent relaxations of these configurations.
%%%%%%%%%%%%%%%%%%%%%%%%%%%%%%%%%%%%

We developed a number of GRACE models with one- (1L) and two-layer (2L) architectures, systematically varying their complexity through small (no suffix), medium (-M suffix), and large (-L suffix) setups. 
The one-layer models are built on ACE star-graphs with direct interactions, the two-layer models include semi-local interactions mediated by equivariant message passing. Both, one-layer and two-layer models employ chemical embedding for efficiently condensing chemical interactions into low rank representations.
The initial parameterizations, designated as ``-OMAT-base'', were conducted using the OMat24 dataset and employed a loss function that equally weighted energies, forces, and stresses.
Further fine-tuning, which used larger weights for the energy loss component, resulted in a series of models designated with an ``-OMAT-ft-E'' or just ``-OMAT'' suffix.

While OMat24 provides a robust foundation, its DFT and pseudopotential settings differ from those used in Alexandria and the Materials Project. To address this, we fine-tuned the OMAT-base models using a combined dataset of MPTraj and  a subsampled Alexandria (sAlex) dataset~\cite{schmidt2023machine,wang2023symmetry,barroso_omat24}. 
The sAlex subset was curated to prevent data leakage with the WBM test set~\cite{wang2021predicting}, a crucial step to ensure model compatibility with Matbench Discovery~\cite{riebesell2025framework}.
The resulting models are denoted with suffix ``-OAM''. 
These GRACE models are designed to serve as robust foundational interatomic potentials and to provide uniform accuracy across a broad range of chemical compositions and structural configurations. 
More details on the models architecture and parameterization can be found in Supplementary Information~\ref{SI:arch_and_param}.

\subsection{Validation}
Foundational MLIPs must demonstrate {uniform accuracy across multiple application domains}. In this section, we present several critical validation tests.
We evaluate our models against the MatBench Discovery benchmark~\cite{riebesell2025framework} for formation energies and thermodynamic stability and the $\kappa$-SRME~\cite{pota2024thermal} test for thermal conductivity, which reflects second- and third-order derivatives.
We further determine the performance of our models on elastic properties and for non-equilibrium and defective configurations by predicting formation energies of grain boundaries, surfaces, and point defects in pure elements. This suite of tests provides a good assessment of the models' capabilities and limitations.

\begin{figure*}[t!]
\centering
\includegraphics[width=\linewidth]{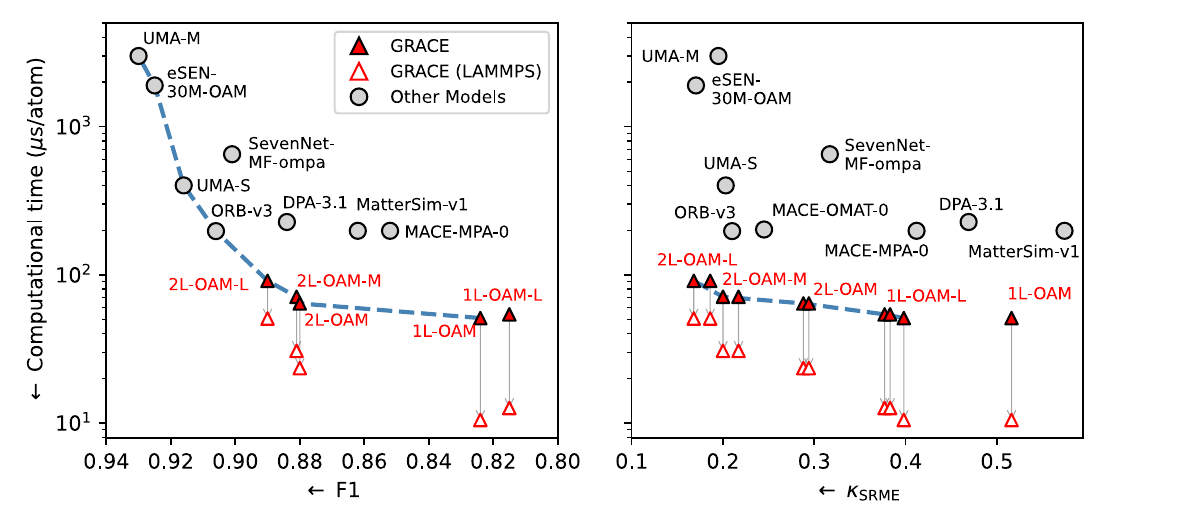}
\caption{
Model performance for stable structure identification (F1 score in MatBench Discovery benchmark) and thermal conductivity prediction ($\kappa_\mathrm{SRME}$) versus computational time per atom. 
A higher F1 score and lower $\kappa_\mathrm{SRME}$ indicate better performance.
The blue dashed line links Pareto optimal models.
Computational performance is estimated via ASE (filled symbols) and LAMMPS (open symbols), with GRACE models indicated in red. 
}
\label{fig:matbench_pareto}
\end{figure*}

\begin{table}[]
    \caption{Performance of foundational MLIPs in materials discovery task (F1), thermal conductivity task ($\kappa_\mathrm{SRME}$) and computational performance (in $\mu$s/atom/step) for W-BCC crystal with 1024 atoms on A100-80GB GPU in ASE and LAMMPS molecular dynamics simulations. 
    Top part of the table corresponds to models compliant with MatBench Discovery task, bottom part to models trained on OMat24 dataset only.}
    \label{tab:f1_kappa_srme_table}
    \centering
\begin{tabular}{|l|c|c|c|c|}
\hline
{Model} & {F1} & {$\kappa_\mathrm{SRME}$} & {t$_\mathrm{ASE}$} & {t$_\mathrm{LAMMPS}$} \\
\hline
UMA-M-1.1 & \textbf{0.930} & 0.195 & 2981 &  \\
eSEN-30M-OAM & 0.925 & 0.170 & 1897 & \\
eqV2 M & 0.917 & 1.771 & -  & \\
UMA-S-1.1 & 0.916 & 0.203 & 401  &  \\
ORB-v3 & 0.906 & 0.210 & 197  & \\
SevenNet-MF-ompa & 0.901 & 0.317 & 651  &  \\
\textbf{GRACE-2L-OAM-L} & 0.890 & \textbf{0.168} & 91  &  51 \\
DPA-3.1-3M-FT & 0.884 & 0.469 & 227  & \\
\textbf{GRACE-2L-OAM-M} & 0.881 & 0.200 & 71  & 31 \\
\textbf{GRACE-2L-OAM} & 0.880 & 0.294 & 64  &  23\\
MatterSim-v1 & 0.862 & 0.574 & 198  & \\
MACE-MPA-0 & 0.852 & 0.412 & 198  & \\
GNOME & 0.829 & - & -  & \\
\textbf{GRACE-1L-OAM} & 0.824 & 0.516 & \textbf{51}  &  10 \\
\textbf{GRACE-1L-OAM-L} & 0.815 & 0.377 & 54  &  13\\
\hline
\textbf{GRACE-2L-OMAT-L} & - & 0.186 & 91  &  51\\
\textbf{GRACE-2L-OMAT-M} & - & 0.217 & 71  &  31\\
MACE-OMAT-0 & - & 0.245 & 202 & \\
\textbf{GRACE-2L-OMAT} & - & 0.288 & 64  &  23 \\
\textbf{GRACE-1L-OMAT-L} & - & 0.383 & 54  &  13\\
\textbf{GRACE-1L-OMAT} & - & 0.398 & \textbf{51}   &  10 \\
\hline
\end{tabular}

\end{table}

\subsubsection{MatBench Discovery benchmark}
Matbench Discovery~\cite{riebesell2025framework} serves as a benchmark for high-throughput discovery of stable inorganic crystals. 
It is specifically designed to evaluate the efficacy of various foundational MLIPs in predicting formation energy and thermodynamic stability of novel crystal structures.
The benchmark task involves geometry optimization of structures sourced from the WBM dataset, a diverse collection of 257,000 candidate crystal structures spanning a wide range of compositions. 
The central goal is to predict formation energies and assess the stability of these structures relative to the original convex hull from Materials Project~\cite{jain2013commentary} with high fidelity. 
Achieving this requires models to accurately predict potential energy, atomic forces, and stress tensors, with an accuracy comparable to the Materials Project's density functional theory (DFT) calculations.

Figure~\ref{fig:matbench_pareto} (left panel) illustrates the performance of GRACE models, specifically the OAM-fine-tuned versions, against other publicly available foundational MLIPs in relation to their computational efficiency. 
The figure shows that GRACE models consistently occupy the Pareto optimality front, demonstrating a superior balance of performance and computational speed. 
This efficiency can be further enhanced by using the LAMMPS molecular dynamics code~\cite{LAMMPS}, with specific timings available in Table~\ref{tab:f1_kappa_srme_table}. 
While the F1 metric reflects a model's accuracy for identifying stable structures, the mean absolute error (MAE) of formation energies offers a more general measure of accuracy. 
The performance of the models for formation energy MAE, which shows a similar trend with GRACE models on the Pareto front, is provided in the supplementary materials.

\subsubsection{Thermal conductivity}

To evaluate the ability of foundational MLIPs to predict force-dependent properties like phonons and anharmonic thermal conductivity, we used the symmetric relative mean error $\kappa_\mathrm{SRME}$ metric~\cite{pota2024thermal}.
This test quantifies a model's performance by predicting the thermal conductivity $\kappa$ across 103 binary structures. 
The thermal conductivity values are calculated from forces predicted by the foundational MLIPs and subsequently analyzed using the \texttt{phono3py} software~\cite{togo2015distributions,togo2023implementation}. 
Accurate thermal conductivity predictions serve as a strong indicator of a model's performance for other simulation tasks, such as modeling metal-organic frameworks~\cite{krass2025mofsimbench}.

The results for various foundational MLIPs, including our GRACE models, are presented in Fig.~\ref{fig:matbench_pareto} (right panel) and Table~\ref{tab:f1_kappa_srme_table}. The family of one- and two-layer GRACE models notably form the Pareto front, achieving the best performance in thermal conductivity prediction. In particular, the GRACE-2L-OAM-L model achieved the lowest error with $\kappa_\mathrm{SRME}=0.168$, underscoring its exceptional accuracy in this domain.

\subsubsection{Elastic properties}

\begin{figure}[h!]
\includegraphics[width=\linewidth]{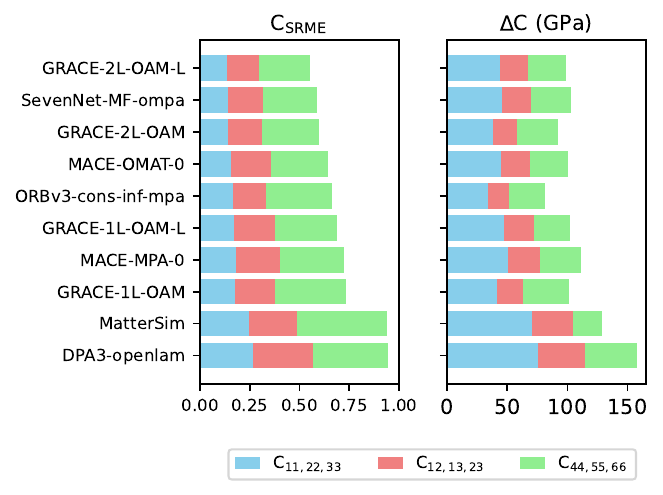}
\caption{The symmetric relative mean error (SRME) and MAE ($\Delta C$ in GPa) for elastic constants, categorized into three subgroups:  longitudinal ($C_{11}, C_{22}, C_{33}$), Poisson's ratio-related ($C_{12}, C_{13}, C_{23}$), and shear ($C_{44}, C_{55}, C_{66}$). See text for more details.}
\label{fig:elast}
\end{figure}

Predicting elastic moduli is a crucial validation test for interatomic potentials.
We categorize elastic constants into three subgroups: longitudinal ($C_{11}, C_{22}, C_{33}$), Poisson's ratio-related ($C_{12}, C_{13}, C_{23}$), and shear ($C_{44}, C_{55}, C_{66}$). 
Because these groups often have varying magnitudes, we primarily focused on the symmetric relative mean error (SRME) and MAE within each subgroup. 
Figure~\ref{fig:elast} presents both metrics with respect to reference data from the Materials Project~\cite{de2015charting}.

Among all tested models, GRACE-2L-OAM-L demonstrated the lowest $C_\mathrm{SRME}$.
Generally, most models showed comparable performance, with the notable exceptions of MatterSim and DPA3-openlam, whose training sets differed from the Materials Project DFT settings used for the reference elastic constant calculations.
A consistent trend across all models is that longitudinal constants typically exhibit lower SRME but higher absolute errors ($\Delta C$) due to their larger magnitudes. In contrast, the second and third groups of elastic constants tend to show smaller absolute errors but larger relative errors.

\subsubsection{Grain boundaries}

\begin{figure}[h]
\includegraphics[width=\linewidth]{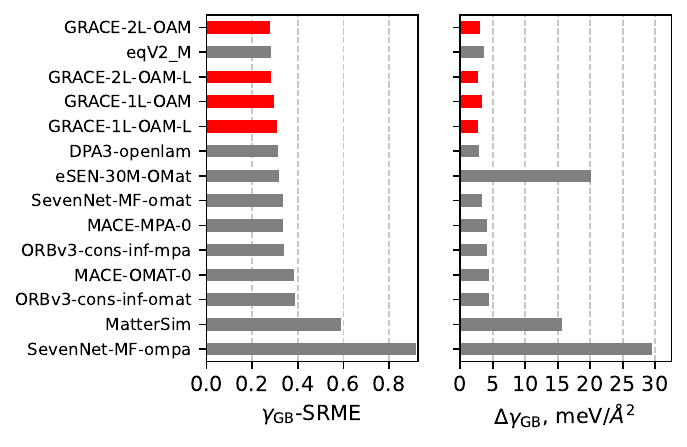}
\caption{Error metrics for unary grain boundaries formation energies: $\gamma_\mathrm{GB}$-SRME (left) and mean absolute error $\Delta \gamma_\mathrm{GB}$ (right). GRACE models are highlighted in red. % \RD{Please also highlight GRACE in red in the right panel.}
}
\label{fig:GBs}
\end{figure}

To assess the performance of foundational MLIPs for bulk structural defects, we utilized an existing dataset of grain boundary formation energies ($\gamma_\mathrm{GB}$ computed for pure metals~\cite{zheng2020grain}. 
The models' accuracy was quantified by calculating both the $\gamma_\mathrm{GB}$-SRME and the mean absolute error $\Delta \gamma_\mathrm{GB}$. The results are presented in Fig.~\ref{fig:GBs}.
 
The relative error $\gamma_\mathrm{GB}$-SRME generally ranges from 0.275 to 0.4, with MatterSim and SevenNet-MF-ompa being notable exceptions. 
A larger $\gamma_\mathrm{GB}$-SRME for K, Rb, and Cs was consistently observed across almost all models, suggesting that the typically used 6~\AA{} cutoff may be insufficient for these alkali elements.
For most models, the absolute error $\Delta \gamma_\mathrm{GB}$ remains below 5~meV/\AA$^2$, with a few exceptions observed for eSEN-30M-Omat, MatterSim, and SevenNet-MF-ompa. More detailed information can be found in the Supplementary Information.

\subsubsection{Surface energies}

\begin{figure}[h]
\includegraphics[width=\linewidth]{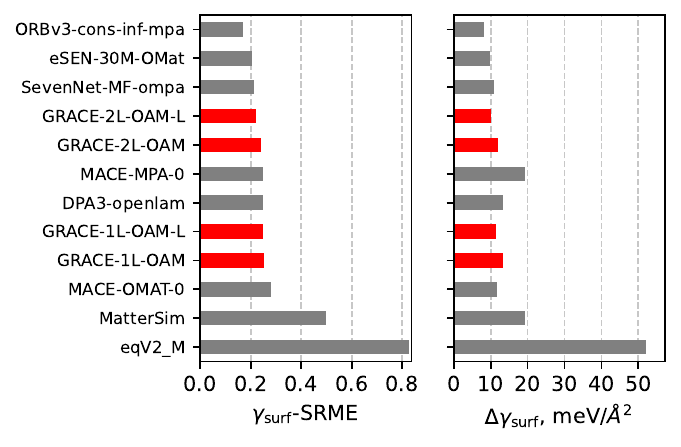}
\caption{Error metrics for unary surface formation energies: $\gamma_\mathrm{surf}$-SRME (left) and mean absolute error $\Delta \gamma_\mathrm{surf}$ (right). GRACE models are highlighted in red.
}
\label{fig:surfs}
\end{figure}

We assessed the ability of foundational MLIPs to predict open structures, such as surfaces, using surface energies ($\gamma_{\mathrm{surf}}$) for pure elements~\cite{tran2016surface}. 
Model accuracy was quantified by calculating both the symmetric relative mean error ($\gamma_{\mathrm{surf}}$-SRME) and the mean absolute error ($\Delta\gamma_{\mathrm{surf}}$), with results presented in Fig.~\ref{fig:surfs}.
Here, the relative error $\gamma_{\mathrm{surf}}$-SRME ranged from 0.168 for the ORBv3-cons-inf-mpa model to 0.279 for the MACE-OMAT-0 model. 
The absolute error $\Delta\gamma_{\mathrm{surf}}$ typically varied from 8 to 14 meV/\AA$^2$, with a few exceptions noted for MACE-MPA-0, MatterSim, and eqV2\_M. 
A consistent finding across all models was poor $\gamma_{\mathrm{surf}}$-SRME metrics for Potassium (K), Rubidium (Rb), Cesium (Cs), and Indium (In).
More detailed information can be found in the Supplementary Information.

\subsubsection{Point defects}
We used systematically computed formation energies for self-interstitials (SIA) and vacancies in BCC~\cite{ma2019universality} and FCC~\cite{ma2021nonuniversal} metals available in the literature as references. 
Given the varying scales of these formation energies across different defect types and metals, we used the symmetric relative mean error metrics, E$_\mathrm{SIA}$-SRME and E$_\mathrm{vac}$-SRME, as main measure of accuracy.

\begin{figure}[h]
\includegraphics[width=\linewidth]{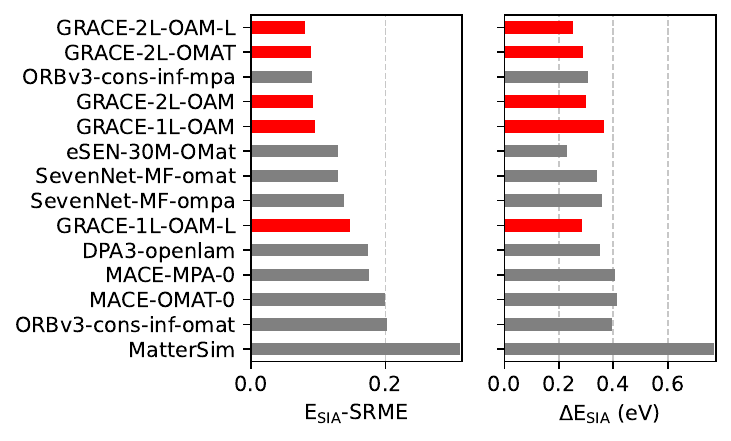}
\includegraphics[width=\linewidth]{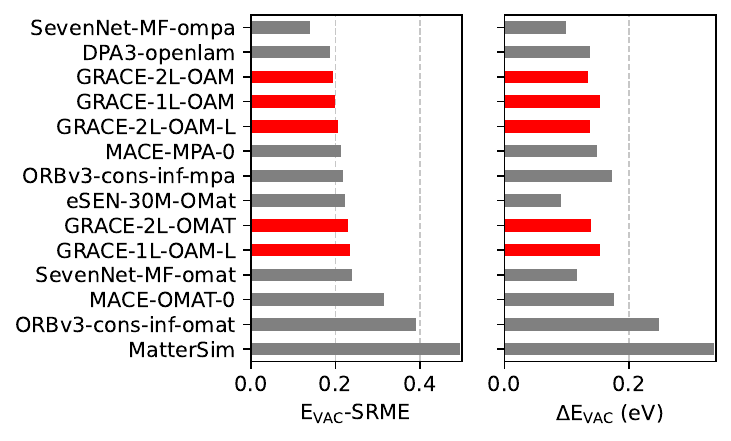}
\caption{Error metrics for SRME of point defects formation energies in unaries: self-interstitials and vacancies. GRACE models are highlighted in red. 
}
\label{fig:SIA}
\end{figure}

As shown in Fig.~\ref{fig:SIA}, the SRME metrics for both defect types generally fall within 0.1 to 0.3, with a few outliers such as MatterSim. 
This discrepancy is likely due to the different DFT settings between MatterSim's training set and the reference data, as observed in our previous analyses. 
In terms of absolute values, the mean absolute error (MAE) for SIA formation energies typically ranges from about 0.2 to 0.4 eV, while for vacancy formation energies, the errors are between 0.1 to 0.2 eV. Further details can be found in the Supplementary Information.

\subsection{Long-time stability of MD}

\begin{figure}
\includegraphics[width=\linewidth]{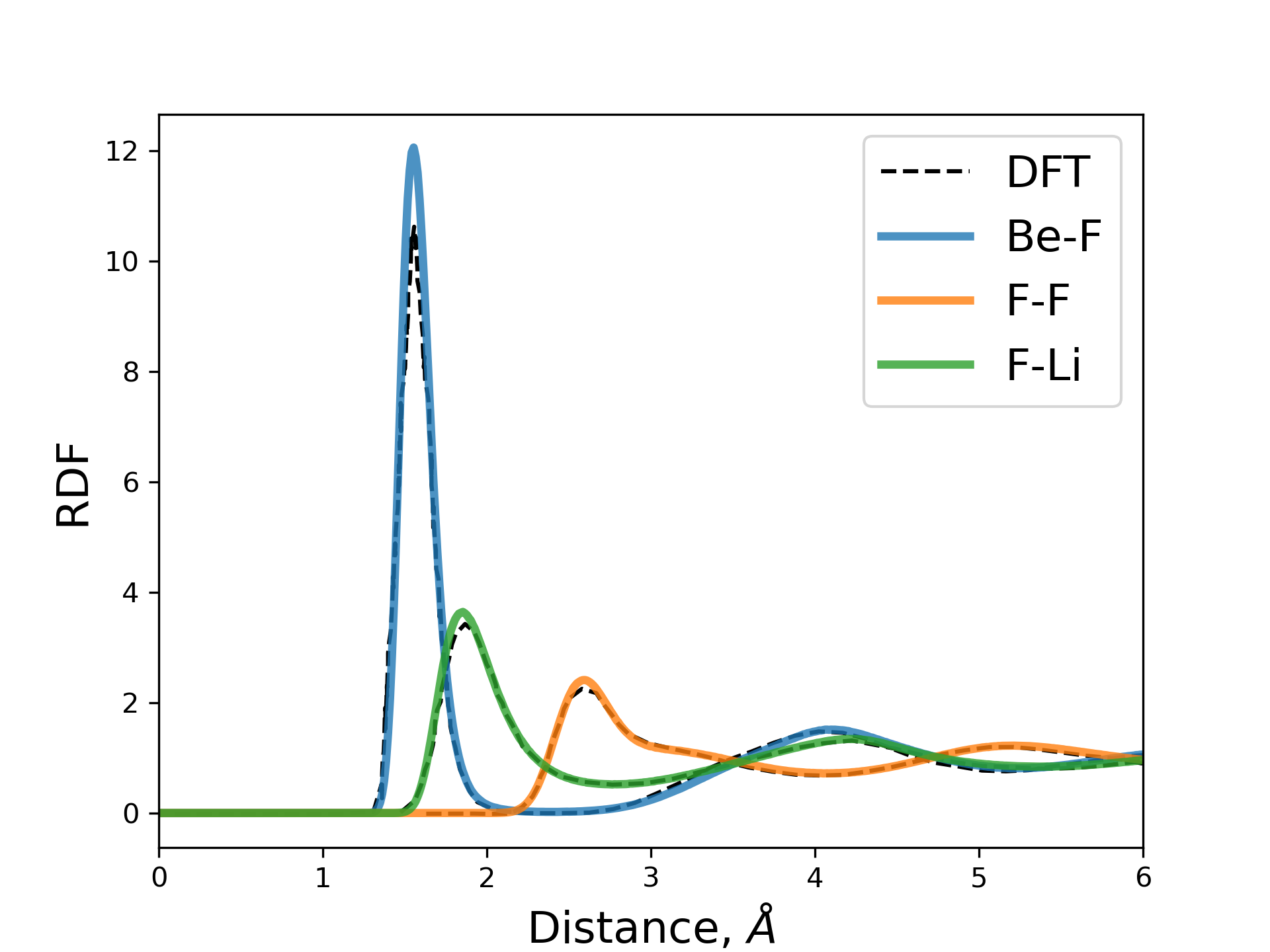}
\caption{RDF of FLiBe salt from MD at 973K using GRACE in comparison to AIMD-DFT results~\cite{lam2021modeling,rodriguez2021thermodynamic}.}
\label{fig:rdf}
\end{figure}

Long-time stability in molecular dynamics (MD) simulations is essential for accurately capturing the dynamical behavior and thermodynamic properties that cannot be fully assessed by static property benchmarks or short simulations.
To demonstrate the out-of-the-box MD stability and performance of the GRACE foundational MLIPs, we conducted a 1\,ns MD simulation of a FLiBe cell containing approximately three thousand atoms at 973\,K, utilizing the GRACE-2L-OMAT-L model \changed{in the NVE ensemble, observing a negligible total energy drift of 5$\cdot10^{-9}$~eV/atom/ns (see Supplementary Information for more details).}We compared the resulting radial distribution functions (RDFs) to reference AIMD data from Ref.~\cite{lam2021modeling,rodriguez2021thermodynamic}, as shown in Fig.~\ref{fig:rdf}. 
Additionally, we estimated diffusion coefficients for each element from our MD simulation, obtaining values of 1.33, 1.58, and $5.86 \times 10^{-5}$~cm$^2$s$^{-1}$ for Be, F, and Li, respectively.
These values align well with the AIMD results ($0.83\pm0.1$, $1.73 \pm 0.17$ and $5.67 \pm 0.52 \times 10^{-5}$~cm$^2$s$^{-1}$, respectively~\cite{lam2021modeling}).
This simulation confirms the model's stability and its accuracy in predicting both structural and dynamical properties over extended timescales.

\subsection{Computational performance}

Computational performance is a critical factor, especially for high-throughput calculations and large-scale, long-duration MD simulations. We evaluated the performance of the GRACE models using LAMMPS on an NVIDIA A100 GPU with 80 GB of memory. Test systems included carbon diamond, liquid water, aluminum FCC, and the molten salt FLiBe, which have different densities and numbers of atomic neighbors. All systems were simulated for a few steps in the NVT ensemble at 300 K, except for FLiBe, which was at 823 K.

\begin{figure}[h]
\includegraphics[width=\linewidth]{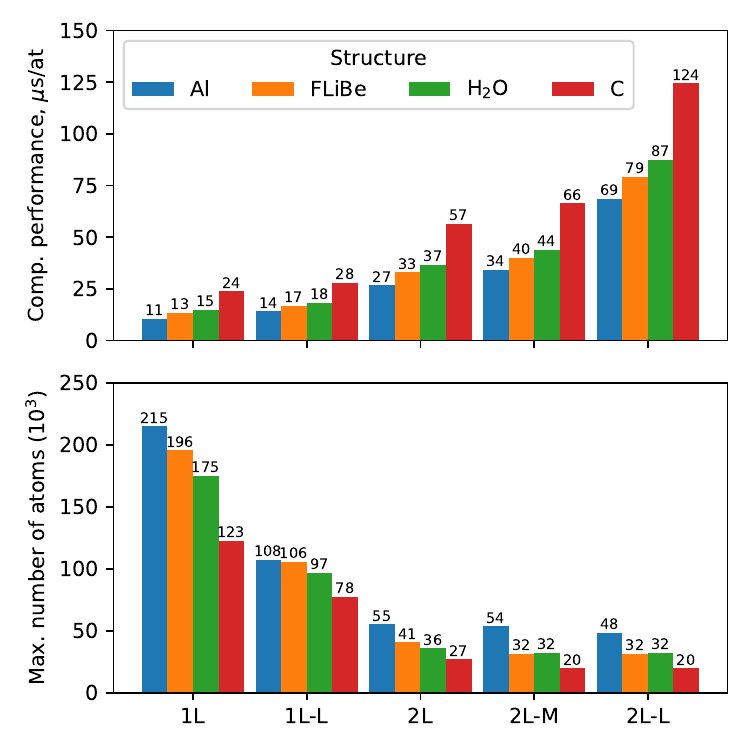}
\caption{Computational performance and memory scaling (maximum number of atoms) of foundational GRACE potentials in LAMMPS on a single A100-80GB GPU, evaluated across diverse materials systems. }
\label{fig:lammps_perf}
\end{figure}

To ensure accurate performance analysis, we excluded the initial MD step, which was significantly slower due to just-in-time (JIT) model compilation. System sizes were increased incrementally until an out-of-memory error occurred. The computational performance, expressed in microseconds per atom, and the maximum number of atoms fitting into memory are presented in Fig.~\ref{fig:lammps_perf}.

The carbon diamond system, with the highest number of neighboring atoms within the cutoff radius, proved to be the most computationally demanding. Still, even for our most intensive model, GRACE-2L-L, up to 20,000 carbon atoms could be accommodated with a performance of approximately 124 µs per atom per step. Across different systems, two-layer GRACE models exhibited computational performance ranging from 27 to 120 µs/atom, while single-layer models ranged from 10 to 28 µs/atom, enabling efficient MD simulations.

Regarding memory usage, a single A100-80GB GPU could accommodate between 20,000 and 55,000 atoms for two-layer GRACE models and 78,000 to 215,000 atoms for one-layer GRACE models. Since the one-layer GRACE model is local (interactions are limited to a cutoff radius), it can be parallelized via domain decomposition as implemented in LAMMPS, which further boosts computational performance for very large systems with millions or billions of atoms.

We also measured the computational performance of other foundational MLIPs by running ten MD steps in ASE~\cite{ase-paper} of W-BCC supercells with different numbers of atoms. These tests utilized different GPU hardware, including the commodity RTX3060 (12 GB), and L40s (40 GB), A100 (80 GB), and H200 (141 GB). The resulting execution times are presented in Table~\ref{tab:MD_perf_ASE}. GRACE models deliver excellent performance even on commodity GPUs like the RTX 3060, despite consistently operating with FP64 (double-precision) accuracy.
\changed{As shown in Fig.\ref{fig:matbench_pareto}, the use of ASE can introduce substantial computational overhead, stemming primarily from Python's execution speed and suboptimal neighbor list construction algorithms.}

\begin{table*}[htbp]  
\centering
\setlength{\tabcolsep}{0pt}
\renewcommand{\arraystretch}{1.25} % Increase row height by 50%

    \caption{Computational performance of GRACE foundational potentials and other foundational MLIPs for MD simulations of a tungsten-BCC supercell in ASE, reported in $\mu$s/atom/step across different GPU architectures.}
    \label{tab:MD_perf_ASE}

\begin{tabular}{|c|c|c|c|c|c|c|c|c|c|c|c|}
\hline
 GPU & \multicolumn{2}{c|}{ RTX 3060 (12 GB) } & \multicolumn{3}{c|}{ L40s (40Gb) } & \multicolumn{3}{c|}{ A100 (80Gb) } & \multicolumn{3}{c|}{ H200 (141Gb) } \\
 \hline %\cline{2-12}
 Num. of atoms & 256 & 512 & 512 & 1k & 2k & 512 & 1k & 2k & 512 & 1k & 2k \\
\hline
GRACE-1L & {\cellcolor[HTML]{E7E419}} \color[HTML]{000000} 111 & {\cellcolor[HTML]{EAE51A}} \color[HTML]{000000} 103 & {\cellcolor[HTML]{EFE51C}} \color[HTML]{000000} 82 & {\cellcolor[HTML]{F1E51D}} \color[HTML]{000000} 70 & {\cellcolor[HTML]{F1E51D}} \color[HTML]{000000} 68 & {\cellcolor[HTML]{F4E61E}} \color[HTML]{000000} 54 & {\cellcolor[HTML]{F4E61E}} \color[HTML]{000000} 51 & {\cellcolor[HTML]{F4E61E}} \color[HTML]{000000} 48 & {\cellcolor[HTML]{F4E61E}} \color[HTML]{000000} 53 & {\cellcolor[HTML]{F6E620}} \color[HTML]{000000} 44 & {\cellcolor[HTML]{F6E620}} \color[HTML]{000000} 42 \\
GRACE-1L-L & {\cellcolor[HTML]{DDE318}} \color[HTML]{000000} 159 & {\cellcolor[HTML]{DFE318}} \color[HTML]{000000} 142 & {\cellcolor[HTML]{ECE51B}} \color[HTML]{000000} 89 & {\cellcolor[HTML]{EFE51C}} \color[HTML]{000000} 77 & {\cellcolor[HTML]{EFE51C}} \color[HTML]{000000} 79 & {\cellcolor[HTML]{F4E61E}} \color[HTML]{000000} 57 & {\cellcolor[HTML]{F4E61E}} \color[HTML]{000000} 54 & {\cellcolor[HTML]{F4E61E}} \color[HTML]{000000} 51 & {\cellcolor[HTML]{F4E61E}} \color[HTML]{000000} 55 & {\cellcolor[HTML]{F6E620}} \color[HTML]{000000} 46 & {\cellcolor[HTML]{F6E620}} \color[HTML]{000000} 45 \\
GRACE-2L & {\cellcolor[HTML]{D0E11C}} \color[HTML]{000000} 214 & {\cellcolor[HTML]{D2E21B}} \color[HTML]{000000} 200 & {\cellcolor[HTML]{EAE51A}} \color[HTML]{000000} 101 & {\cellcolor[HTML]{ECE51B}} \color[HTML]{000000} 92 & {\cellcolor[HTML]{ECE51B}} \color[HTML]{000000} 90 & {\cellcolor[HTML]{F1E51D}} \color[HTML]{000000} 68 & {\cellcolor[HTML]{F1E51D}} \color[HTML]{000000} 64 & {\cellcolor[HTML]{F1E51D}} \color[HTML]{000000} 62 & {\cellcolor[HTML]{F1E51D}} \color[HTML]{000000} 59 & {\cellcolor[HTML]{F4E61E}} \color[HTML]{000000} 50 & {\cellcolor[HTML]{F4E61E}} \color[HTML]{000000} 49 \\
GRACE-2L-M & {\cellcolor[HTML]{BDDF26}} \color[HTML]{000000} 303 & {\cellcolor[HTML]{C0DF25}} \color[HTML]{000000} 292 & {\cellcolor[HTML]{E5E419}} \color[HTML]{000000} 118 & {\cellcolor[HTML]{E7E419}} \color[HTML]{000000} 109 & {\cellcolor[HTML]{E7E419}} \color[HTML]{000000} 111 & {\cellcolor[HTML]{EFE51C}} \color[HTML]{000000} 74 & {\cellcolor[HTML]{EFE51C}} \color[HTML]{000000} 71 & {\cellcolor[HTML]{F1E51D}} \color[HTML]{000000} 69 & {\cellcolor[HTML]{F1E51D}} \color[HTML]{000000} 63 & {\cellcolor[HTML]{F4E61E}} \color[HTML]{000000} 54 & {\cellcolor[HTML]{F4E61E}} \color[HTML]{000000} 50 \\
GRACE-2L-L & {\cellcolor[HTML]{A5DB36}} \color[HTML]{000000} 405 & {\cellcolor[HTML]{A8DB34}} \color[HTML]{000000} 394 & {\cellcolor[HTML]{DFE318}} \color[HTML]{000000} 142 & {\cellcolor[HTML]{E2E418}} \color[HTML]{000000} 133 & {\cellcolor[HTML]{E2E418}} \color[HTML]{000000} 140 & {\cellcolor[HTML]{EAE51A}} \color[HTML]{000000} 95 & {\cellcolor[HTML]{ECE51B}} \color[HTML]{000000} 91 & {\cellcolor[HTML]{EAE51A}} \color[HTML]{000000} 95 & {\cellcolor[HTML]{EFE51C}} \color[HTML]{000000} 76 & {\cellcolor[HTML]{F1E51D}} \color[HTML]{000000} 66 & {\cellcolor[HTML]{F1E51D}} \color[HTML]{000000} 63 \\
ORBv3(fp32) & {\cellcolor[HTML]{8ED645}} \color[HTML]{000000} 506 & {\cellcolor[HTML]{95D840}} \color[HTML]{000000} 470 & {\cellcolor[HTML]{DFE318}} \color[HTML]{000000} 142 & {\cellcolor[HTML]{E2E418}} \color[HTML]{000000} 129 & {\cellcolor[HTML]{E5E419}} \color[HTML]{000000} 125 & {\cellcolor[HTML]{D2E21B}} \color[HTML]{000000} 208 & {\cellcolor[HTML]{D5E21A}} \color[HTML]{000000} 197 & {\cellcolor[HTML]{D5E21A}} \color[HTML]{000000} 195 & {\cellcolor[HTML]{E7E419}} \color[HTML]{000000} 114 & {\cellcolor[HTML]{EAE51A}} \color[HTML]{000000} 97 & {\cellcolor[HTML]{ECE51B}} \color[HTML]{000000} 89 \\
ORBv3(fp64) & {\cellcolor[HTML]{440154}} \color[HTML]{F1F1F1} 15566 & {\cellcolor[HTML]{440154}} \color[HTML]{F1F1F1} 15328 & {\cellcolor[HTML]{433E85}} \color[HTML]{F1F1F1} 2451 & {\cellcolor[HTML]{414287}} \color[HTML]{F1F1F1} 2407 & {\cellcolor[HTML]{404588}} \color[HTML]{F1F1F1} 2381 & {\cellcolor[HTML]{CDE11D}} \color[HTML]{000000} 231 & {\cellcolor[HTML]{CDE11D}} \color[HTML]{000000} 223 & {\cellcolor[HTML]{D0E11C}} \color[HTML]{000000} 221 & {\cellcolor[HTML]{E2E418}} \color[HTML]{000000} 131 & {\cellcolor[HTML]{E5E419}} \color[HTML]{000000} 119 & {\cellcolor[HTML]{E7E419}} \color[HTML]{000000} 114 \\
MatterSim & {\cellcolor[HTML]{A5DB36}} \color[HTML]{000000} 409 & {\cellcolor[HTML]{B8DE29}} \color[HTML]{000000} 319 & {\cellcolor[HTML]{DFE318}} \color[HTML]{000000} 149 & {\cellcolor[HTML]{E5E419}} \color[HTML]{000000} 124 & {\cellcolor[HTML]{DFE318}} \color[HTML]{000000} 151 & {\cellcolor[HTML]{C0DF25}} \color[HTML]{000000} 288 & {\cellcolor[HTML]{D5E21A}} \color[HTML]{000000} 198 & {\cellcolor[HTML]{DDE318}} \color[HTML]{000000} 163 & {\cellcolor[HTML]{CAE11F}} \color[HTML]{000000} 244 & {\cellcolor[HTML]{DFE318}} \color[HTML]{000000} 145 & {\cellcolor[HTML]{ECE51B}} \color[HTML]{000000} 92 \\
MACE-OMAT-0(cuEq) & {\cellcolor[HTML]{69CD5B}} \color[HTML]{000000} 683 & {\cellcolor[HTML]{73D056}} \color[HTML]{000000} 640 & {\cellcolor[HTML]{D8E219}} \color[HTML]{000000} 177 & {\cellcolor[HTML]{DDE318}} \color[HTML]{000000} 153 & {\cellcolor[HTML]{E2E418}} \color[HTML]{000000} 140 & {\cellcolor[HTML]{C0DF25}} \color[HTML]{000000} 288 & {\cellcolor[HTML]{D2E21B}} \color[HTML]{000000} 202 & {\cellcolor[HTML]{E2E418}} \color[HTML]{000000} 129 & {\cellcolor[HTML]{BDDF26}} \color[HTML]{000000} 296 & {\cellcolor[HTML]{DDE318}} \color[HTML]{000000} 162 & {\cellcolor[HTML]{EAE51A}} \color[HTML]{000000} 99 \\
SevenNet-MF-OMPA & {\cellcolor[HTML]{481F70}} \color[HTML]{F1F1F1} 2744 & {\cellcolor[HTML]{000000}} \color[HTML]{F1F1F1} {\cellcolor{lightgray}} OOM & {\cellcolor[HTML]{1FA187}} \color[HTML]{F1F1F1} 1283 & {\cellcolor[HTML]{1FA187}} \color[HTML]{F1F1F1} 1289 & {\cellcolor[HTML]{000000}} \color[HTML]{F1F1F1} {\cellcolor{lightgray}} OOM & {\cellcolor[HTML]{67CC5C}} \color[HTML]{000000} 700 & {\cellcolor[HTML]{70CF57}} \color[HTML]{000000} 651 & {\cellcolor[HTML]{73D056}} \color[HTML]{000000} 633 & {\cellcolor[HTML]{AADC32}} \color[HTML]{000000} 380 & {\cellcolor[HTML]{B0DD2F}} \color[HTML]{000000} 357 & {\cellcolor[HTML]{B5DE2B}} \color[HTML]{000000} 334 \\
eSEN-30M-OMat & {\cellcolor[HTML]{000000}} \color[HTML]{F1F1F1} {\cellcolor{lightgray}} OOM & {\cellcolor[HTML]{000000}} \color[HTML]{F1F1F1} {\cellcolor{lightgray}} OOM & {\cellcolor[HTML]{355E8D}} \color[HTML]{F1F1F1} 2120 & {\cellcolor[HTML]{000000}} \color[HTML]{F1F1F1} {\cellcolor{lightgray}} OOM & {\cellcolor[HTML]{000000}} \color[HTML]{F1F1F1} {\cellcolor{lightgray}} OOM & {\cellcolor[HTML]{2E6D8E}} \color[HTML]{F1F1F1} 1937 & {\cellcolor[HTML]{2D718E}} \color[HTML]{F1F1F1} 1897 & {\cellcolor[HTML]{000000}} \color[HTML]{F1F1F1} {\cellcolor{lightgray}} OOM & {\cellcolor[HTML]{4EC36B}} \color[HTML]{000000} 834 & {\cellcolor[HTML]{58C765}} \color[HTML]{000000} 779 & {\cellcolor[HTML]{5CC863}} \color[HTML]{000000} 753 \\
\hline
\end{tabular}

\end{table*}

\subsection{Fine-tuning}

Foundational interatomic potentials are trained to provide accurate simulations across the periodic table, but they may lack the specific precision or performance necessary for particular downstream tasks. Fine-tuning serves as an effective strategy to address these limitations by adapting a pre-existing foundational MLIP. This is done by continuing the training on a new, often small and specialized dataset. This process aims to enhance the model's performance or tailor its predictions for specific chemical systems, properties, or higher levels of theoretical accuracy, while leveraging the extensive knowledge and robust representations acquired during its initial foundational training.

Here we demonstrate the fine-tuning of GRACE-2L foundational MLIP for the Al-Li binary system~\cite{menon2024electrons} and a hydrogen combustion dataset~\cite{guan2022benchmark}.

\subsubsection{Al-Li}
For fine-tuning, we utilized a dataset for the Al-Li binary system from Ref.~\cite{menon2024electrons}. We curated the data by removing structures from the liquid phase and those corresponding to randomly sampled space groups. This resulted in a total of three thousand structures within 1\,eV/atom above the convex hull, from which we allocated 5\% as a test set. 
To study data efficiency, we then created a series of training subsets of varying sizes (5\%, 10\%, 25\%, 50\%, and 75\%) from the remaining data.

Using these training and test sets, we fine-tuned the 1L-OMAT and 2L-OMAT models by updating all weights, a process we term "naive fine-tuning." For comparison, we also trained two models from scratch: a one-layer ACE model as implemented in the PACE software~\cite{Lysogorskiy21,Bochkarev22} and a GRACE-2L model, designated 2L-baseline, with a complexity identical to GRACE-2L-OMAT.

\begin{figure*}[t]

\includegraphics[width=\textwidth]{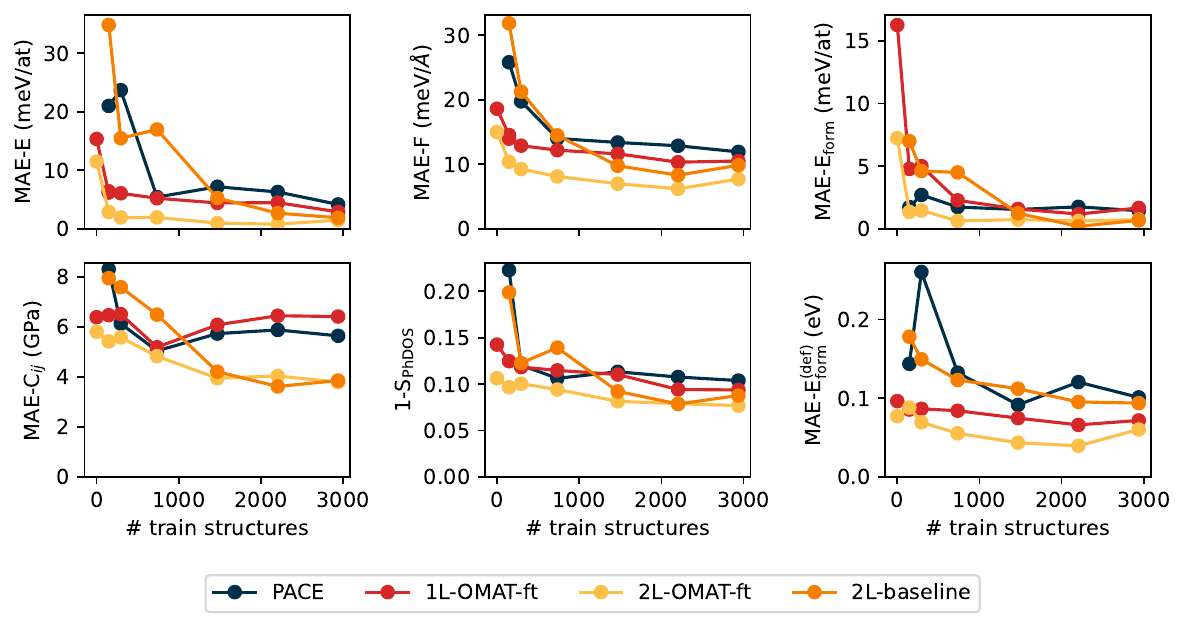}
\caption{Error metrics for fine-tuning on Al-Li dataset. Evaluated properties and corresponding MAE: energies (meV/atom), forces (meV/\AA{}), formation energies ($\mathrm{E}_\mathrm{form}$, meV/atom), elastic matrix elements ($\mathrm{C}_{ij}$, GPa), phonon density of states (PhDOS) and vacancy formation energies ($\mathrm{E}_\mathrm{form}^\mathrm{(def)}$). PhDOS error is measured by Tanimoto similarity $1-\mathrm{S}_\mathrm{PhDOS}$. Data at zero number of train structures correspond to zero-shot models.
}
\label{fig:AlLi-finetun}
\end{figure*}

As shown in Fig.~\ref{fig:AlLi-finetun}, zero-shot predictions without fine-tuning demonstrate very good accuracy for both GRACE-1L-OMAT and GRACE-2L-OMAT. The fine-tuned 2L-OMAT-ft model consistently outperforms the other models, even with small fractions of the curated dataset, while the 2L-baseline model only reaches comparable accuracy with more data. The fine-tuned 1L-OMAT-ft model is slightly less accurate than the two-layer models for most tests and shows comparative performance to the specialized PACE model. We attribute this good performance to the small number of elements and the relatively limited Al-Li dataset, which primarily includes close-to-equilibrium structures.

These results demonstrate that fine-tuning foundational GRACE potentials can be superior to training from scratch, especially in low-data regimes.

\subsubsection{Hydrogen combustion}

%%%%%%%%%%%%%%%%%%%%%%%%%%%%%%%%%%%%%%%%%%%%%%%%%5
The hydrogen combustion (H2COMB) dataset~\cite{guan2022benchmark} includes intrinsic reaction coordinate (IRC) calculations, \textit{ab initio} MD simulations, and normal mode displacement calculations, covering 19 reaction channels for hydrogen combustion. 
This dataset was computed using Q-Chem with $\omega$B97X-V/cc-pVTZ. 
These DFT settings differ significantly from those of the OMat24, Alexandria, and MPTraj datasets that we used to train the foundational GRACE OMAT potentials. 
Consequently, zero-shot predictions by all foundational models show a rather high error. 
For the GRACE-2L-OAM model, the force MAE is 740.5 meV/\AA~\cite{peng2025lambench, openlam_benchmark}.

Due to the systematic difference in datasets, naive fine-tuning may result in catastrophic forgetting - the tendency to forget previously learned information when learning a new task. 
To fine-tune a GRACE-2L-OAM model to the H2COMB dataset without this issue for other elements, we explored several strategies:
1) naive fine-tuning: all parameters are trainable; 
2) frozen-weights approach: only ACE expansion coefficients are trainable, while all other parameters (including chemical embeddings, radial functions, and the energy readout) are kept unchanged. 
Within this approach, we considered two cases (a) only the coefficients of the final ACE expansions before atomic energy readout are trainable and (b) the coefficients of the final ACE expansions and the ACE expansion coefficients for messages passed between the first and second layers are trainable.
These coefficients depend on the central atom type; thus, parameters for elements absent in the fine-tuning dataset will not be updated.

\begin{figure}[t]

\includegraphics[width=\columnwidth]{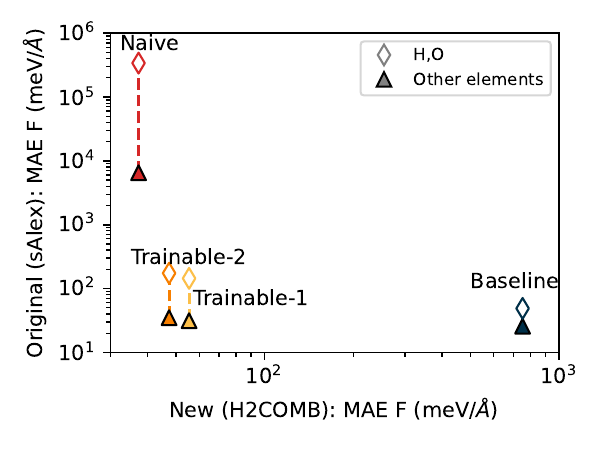}
\caption{Mean Absolute Error (MAE) of forces for fine-tuned GRACE-2L-OAM models. The x-axis shows MAE on the new H2COMB dataset (downstream task), the y-axis shows MAE on the original sAlex dataset. Performance for sAlex is split into H,O atoms (light grey diamonds) and other elements (dark grey triangles). 'Baseline' refers to the zero-shot model. 'Naive' corresponds to strategy 1 (naive fine-tuning). 'Trainable-1' and 'Trainable-2' correspond to strategies 2a and 2b, respectively, representing different frozen-weights approaches.}
\label{fig:HCOMB-finetune}
\end{figure}

The results of these strategies are shown in Figure~\ref{fig:HCOMB-finetune}. 
The baseline zero-shot model shows a very high force MAE for the H2COMB dataset but a low error on the original sAlex dataset. The latter's performance is computed on two subsets: H and O atoms, and all other elements. 
Naive fine-tuning (strategy 1) achieved the best error metrics on the downstream task (37 meV/\AA), but the error on the original dataset increased drastically to 339 $\times 10^{3}$ meV/\AA for H/O and 6.5 $\times 10^{3}$ meV/\AA for the remaining elements, confirming catastrophic forgetting for all elements.

In contrast, the frozen-weights approaches show only a small increase in force MAE on the original sAlex dataset for elements excluding H and O (from 25 meV/\AA for the baseline model to 31 meV/\AA and 35 meV/\AA for strategies 2a and 2b, respectively).
The predictions of these models remain unchanged from the baseline if the structures contain neither H nor O atoms, indicating that the increase in metrics is associated with the presence of H/O atoms within the cutoff. 
For H and O atoms from the sAlex dataset, errors increase much less than with naive fine-tuning (from 49 meV/\AA for the baseline to 144 meV/\AA and 174 meV/\AA), while downstream task errors became 55 and 47 meV/\AA for strategies 2a and 2b, respectively.

The errors on the original versus the new dataset form a Pareto front as displayed in Fig.~\ref{fig:HCOMB-finetune}). By freezing certain parts of the model, catastrophic forgetting of the original task can be mitigated, even for the elements that are updated and despite a systematic shift in reference data. Other methods for fine-tuning foundational GRACE potentials that can mitigate catastrophic forgetting, such as low-rank adaptation (LoRA) or delta tuning, will be considered in future work.

%%%%%%%%%%%%%%%%%%%%%%%%%%%%%%%%%%%%%%%%%%%%%%%%%%

\subsection{Model distillation}

\begin{figure}[t]
\includegraphics[width=\columnwidth]{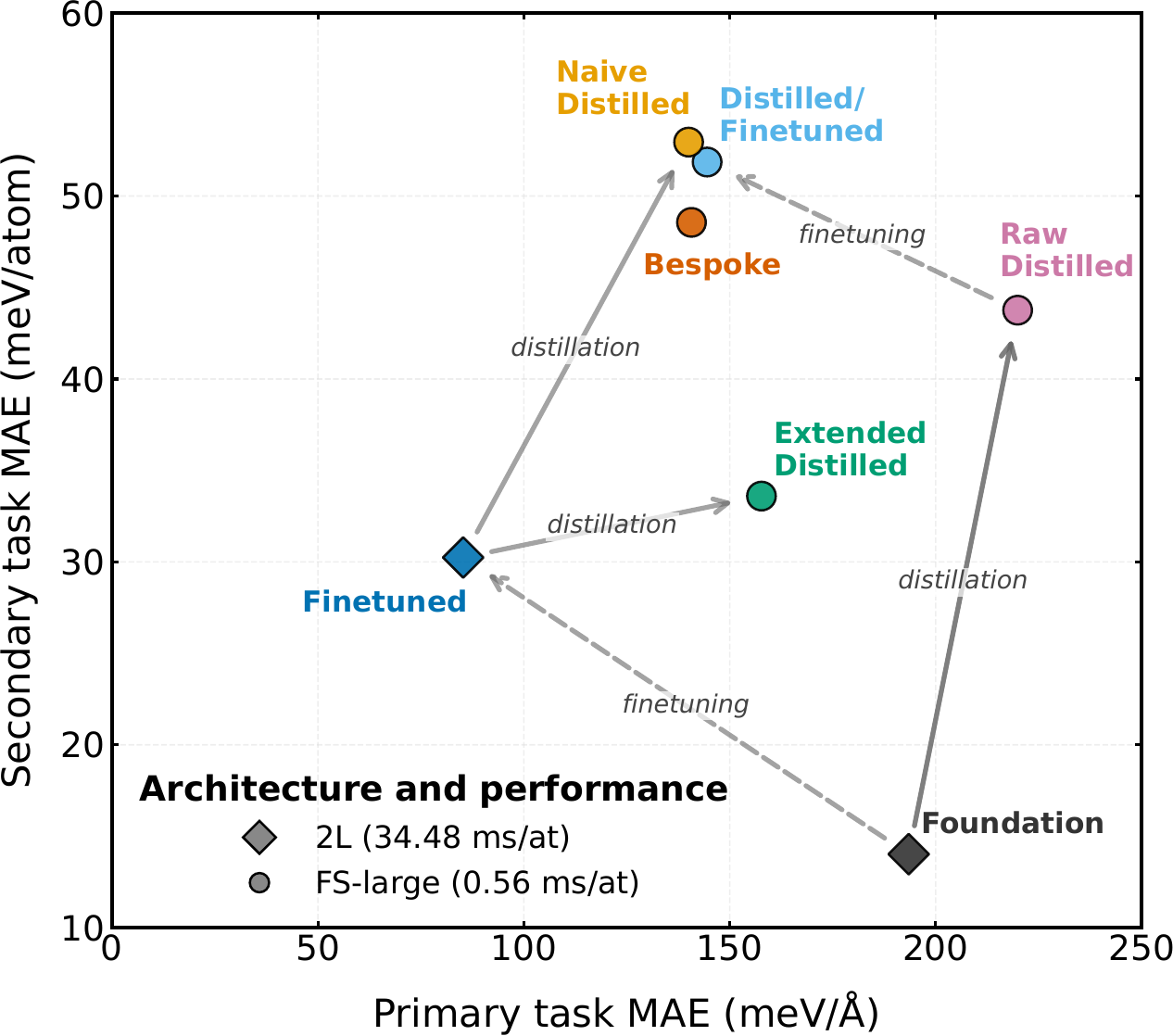}
\caption{\changed{Comparison of distillation and fine-tuning strategies. 
The axes illustrate the trade-off between accuracy on the primary task (HEA25S force MAE) and general chemical stability (secondary task: formation energy MAE of non-magnetic binaries from the Materials Project). 
Computational performance for both GRACE-2L and GRACE-FS architectures was evaluated on a 10-core CPU and normalized per core. See text for details.}}
\label{fig:HEA25-distill-finetune}
\end{figure}

While foundational MLIPs offer high accuracy across a vast chemical and configurational space, they often have lower computational performance and limited parallelization compared to chemistry-specific models. 
To bridge this gap, model distillation can be employed~\cite{morrow2022indirect,gardner2025distillation}, a process involving the retraining of a simpler, faster ``student'' model on a dataset labeled by a foundational ``teacher'' model.
\changed{
We investigate different distillation and fine-tuning pathways using the combined HEA25~\cite{lopanitsyna2023modeling} and HEA25S~\cite{mazitov2024surface} datasets as a case study. We employ the foundational GRACE-2L-OMAT model as the ``teacher'' and the GRACE-FS architecture as the ``student'', selected for its straightforward parallelization and CPU-only inference capabilities.

We evaluated performance based on two tasks with corresponding accuracy metrics. 
The primary task reflects model accuracy on the new downstream application, measured by the force component MAE on the HEA25S validation set. 
The secondary task serves as a proxy for the model's retention of general chemical knowledge and stability, as the HEA25S dataset lacks pure unary or binary structures.
This was measured by the MAE of formation energy of binaries compounds from the Materials Project comprised of non-magnetic elements covered in the HEA25 dataset.

The different pathways for the downstream task are illustrated in Fig.~\ref{fig:HEA25-distill-finetune}. 
The initial foundation model (``Foundation'') exhibits a large error on the primary task due to the DFT functional mismatch between its training data (PBE) and the HEA25S target (PBEsol). 
However, it retains a low error (14\,meV/atom) on the secondary task—comparable to the 23\,meV/atom formation energy MAE on the Matbench Discovery leaderboard.

We explored three distinct approaches to address this. 
In the first method,  we fine-tuned the foundation model on the HEA25S dataset to create a new teacher (``Finetuned''). 
This teacher, having achieved the lowest error on the primary task, was then used to re-label the HEA25S dataset for parameterizing the GRACE-FS  student model called ``Naive Distilled''. 
A second pathway is opposite to the first: we re-labeled the HEA25S dataset using the original foundation model (GRACE-2L-OMAT) to parameterize a ``Raw distilled'' student model, which was subsequently fine-tuned using the original HEA25S dataset to produce the ``Distilled/Finetuned'' model. 
Finally, the third approach involved parameterizing the GRACE-FS model from scratch using the HEA25S dataset directly to create the ``Bespoke'' model. 
As shown in Fig.~\ref{fig:HEA25-distill-finetune}, all three approaches yield models with similar error metrics, though the bespoke model performs slightly better on the primary task due to the usage of the original DFT data.

To improve performance on the secondary task, while preserving good accuracy of the primary task, we generated a synthetic training dataset comprising structures from HEA25 and HEA25S, along with unary and binary structures (including both ideal and rattled configurations) for all 25 elements. 
This extended dataset was labeled by the GRACE-2L teacher (``Finetuned'') and used to parameterize a new student model (``Extended Distilled'').
This model demonstrates significantly improved performance on the secondary task, close to the teacher model, while retaining metrics on the primary task that are only slightly worse to the bespoke model.

We consistently observe a slight degradation in accuracy for student models compared to their teachers due to the simpler architecture of the former.
However, this simplicity yields substantial gains in computational efficiency. 
On a CPU with 10 physical cores, the GRACE-2L model requires approximately 34.48\,ms/atom/core. 
In contrast, the GRACE-FS model, using a standalone C++ implementation, achieves 0.56\,ms/atom/core, representing a speedup of nearly 70$\times$. Thus, the extended distillation strategy offers the optimal balance, recovering the generalizability lost in standard fine-tuning or bespoke training.
The effect of different complexities of teacher and student models, together with other details are presented in the Supplementary Information.
}

\section{Discussion}

We present a series of foundational machine learning interatomic potentials (MLIPs) based on the Graph Atomic Cluster Expansion (GRACE). 
GRACE builds on a formally complete mathematical basis of the local and semi-local atomic environment. Our foundational GRACE potentials obey fundamental physical symmetries and guarantee invariance under rotation, inversion, translation, as well as permutation of identical atomic species. Forces are conservative and computed as gradients of the energy. We further demonstrate computational efficiency at high precision (FP64). This sets GRACE apart from many other MLIPs that employ uncontrolled basis approximations, replace exact symmetries or exact gradients by numerical estimates, or lower precision to FP32 for computational efficiency.

Our foundational GRACE potentials were trained on a massive dataset of DFT calculations that encompasses a broad range of atomic configurations and chemistries.
Comprehensive validation across diverse simulation tasks demonstrates excellent performance and versatility of the GRACE potentials. On the MatBench Discovery leaderboard, our models largely define the Pareto front, showcasing a superior trade-off between predicted thermodynamic stability, formation energy MAE and computational speed. Furthermore, the two-layer GRACE models exhibit leading performance in predicting thermal conductivity, a property highly sensitive to a model's ability to capture anharmonic contributions to atomic displacements and higher-order derivatives. This highlights the robustness of the GRACE framework beyond simple energy and force predictions.

Further validation on structural defects - grain boundaries, surfaces, and point defects - confirms that the GRACE models are able to describe effectively non-equilibrium and open structures. Some outliers, for example, for the alkali metals K, Rb and Cs, are attributed simply to the cut-off radius, which is too small for these large elements.

Long-time molecular dynamics simulations of FLiBe molten salt demonstrated the GRACE models' ability to maintain stability and accurately capture dynamic properties, such as radial distribution functions and diffusion coefficients, over extended timescales.

Beyond the core validation, we explored the practical application of foundational GRACE potentials through fine-tuning and distillation.
We show that fine-tuning GRACE models on small, specialized datasets for Al-Li significantly improves their accuracy for specific downstream tasks.
This approach is particularly effective in low-data regimes, outperforming models trained from scratch.
Moreover, our investigation into catastrophic forgetting demonstrates that freezing specific model layers can mitigate a model's tendency to lose general knowledge when learning new, distinct tasks, as illustrated for a dataset specific for hydrogen combustion.
Finally, we successfully distilled a complex GRACE model into a simpler, more computationally efficient one, while also improving its accuracy on a wider configurational space in comparison to a model trained from scratch.
This shows that foundational potentials can act as powerful ``teachers'' and opens a path for creating specialized, high-performance ACE and GRACE potentials for specific applications without the need for extensive new DFT calculations.

\section{Methods}

\subsection{Models architecture and parameterization}

All GRACE models were implemented in the grace-tensorpotential package, which is based on the TensorFlow library~\cite{tensorflow}. The models implement a graph extension of the atomic cluster expansion (ACE) method, utilizing both star- and tree-like many-body basis functions, that form an orthonormal and complete basis set~\cite{bochkarev2024}. 
The configurations for the GRACE foundation potentials together with parametrization details are outlined in Supplementary Information~\ref{SI:arch_and_param}.

\subsection{Validation}

\textbf{Foundational machine learning interatomic potentials.}
For the MACE models, the 'medium-mpa-0' checkpoint was used for MACE-MPA-0, and 'mace-omat-0-medium' for MACE-OMAT-0. % mace-torch==0.3.12, cuequivariance                0.3.0
cuequivariance-ops-torch-cu12 0.3.0
cuequivariance-torch          0.3.0
All MACE models were using float64 precision and had the \texttt{cuEQ} option enabled. 
The MatterSim model utilized the 'mattersim-v1.0.0-5m' checkpoint. 
The ORB model was based on the 'v3-conservative-inf' modification, configured for 'float32-highest' precision, that corresponds to full float32 precision. 
SevenNet employed the '7net-mf-ompa' checkpoint with its 'mpa' modality. 
For eSEN, the 'esen\_30m\_omat' checkpoint with a seed of 0 was selected.
UMA-M and UMA-S models corresponded to their 'uma-m-1p1' and 'uma-s-1p1' checkpoints, respectively, specifying the 'omat' task. 
The DPA model used the '2025-01-10-dpa3-openlam' checkpoint, whereas for computational performance tests ``DPA-3.1-3M-ft'' checkpoint was used. 
For eqV2\_M model ``eqV2\_86M\_omat\_mp\_salex'' checkpoint was used. 
All GRACE models consistently operated with float64 precision.

\textbf{Computational performance}. The computational performance of foundational MLIPs, depicted in Fig.~\ref{fig:matbench_pareto}, was determined by measuring the averaged wall time for ten molecular dynamics (MD) steps. These simulations were performed for a 1024-atom  supercell of tungsten BCC using ASE on a single NVIDIA A100 GPU with 80 GB of memory. 
Initial runs were excluded to account for Just-In-Time (JIT) compilation and other caching effects. 
Each MD simulation was independently repeated ten times and the results averaged. 
The final metrics are normalized by the number of atoms and MD steps.

\textbf{Elastic constants.} 
We used elastic properties from the Materials Project~\cite{de2015charting} as our reference, employing an energy-based method~\cite{golesorkhtabar2013elastic} to compute the elastic tensor  $C_{ij}$ in Voigt notation. 
Out of 10073 reference elastic matrix calculations, we could not compute elastic constants for all structures due to relaxation and convergence issues with some potentials. Therefore, we relied on a common subset of 7962 structures for which elastic tensors were successfully computed by all potentials. This subset serves as a robust and representative test set for evaluating elastic matrix predictions. The eSEN model was not validated due to its high computational expenses.

\textbf{Grain boundaries.} 
Grain boundary structures and their corresponding reference structures were sourced from the Crystalium project~\cite{zheng2020grain}, a dataset closely related to the Materials Project~\cite{jain2013commentary}. 
The reference structures were fully relaxed using the BFGS method with a FrechetCellFilter, applying a relaxation criterion of 0.001 eV/\AA. 
Grain boundary relaxation was carried out using the FIRE minimization algorithm, also with a FrechetCellFilter, enforcing a maximum force threshold of 0.01 eV/\AA\,  and a limit of 500 optimization steps.
For eqV2 model, relaxation criteria was loosen to 0.02 eV/\AA\, due to numerical instabilities.
In total, 327 grain boundaries were initially computed across 58 different elements. To ensure consistency and avoid discrepancies, only structures with a grain boundary plane orthogonal to the z-direction were selected, resulting in a final set of 297 grain boundary structures.

\textbf{Surfaces.} 
Surface structures and their corresponding reference data were sourced from the Crystalium project~\cite{tran2016surface}, which is associated with the Materials Project~\cite{jain2013commentary}. 
The same relaxation settings as for grain boundaries were applied. 
From an initial total of 1124 surface structures, only those with a surface plane orthogonal to the z-direction were selected, yielding a final set of 716 surface structures for analysis.

\textbf{Self-interstitials and vacancies.}
% Ref~\cite{medasani2015vacancy}, values computed with PBE functional.
Reference data for self-interstitial and vacancy formation energies were taken from Ref.~\cite{ma2019universality} for 13 BCC metals and Ref.~\cite{ma2021nonuniversal} for 13 FCC metals. All these reference values were computed using the PBE functional.

\section{Data availability}
Training datasets (MPTrj, sAlex and OMat24) are publicly available.
GRACE foundational potentials are available at gracemaker.readthedocs.io/en/latest/gracemaker/foundation

\section{Code availability}
Code for GRACE potential is available at github.com/ICAMS/grace-tensorpotential

\section{Funding}
We acknowledge funding from the Deutsche Forschungsgemeinschaft (DFG, German Research Foundation) through SFB 1394 (project number 409476157).

\section{Acknowledgements}
 High-performance computing (HPC) resources were provided by the Paderborn Center for Parallel Computing (PC$^2$) and the Elysium HPC cluster at Ruhr-University Bochum.

\section{Author contributions}
Conceptualisation and Project Administration: All authors. 
Y.L. and A.B. developed the software and parameterized the models.
Writing - original draft: Y.L. 
Writing-review and editing: All authors.
Resources and funding acquisition: Y.L., R.D.

\section{Competing interests}
The authors declare no competing interests.

\bibliography{biblio_R1.bib}

\clearpage
\section*{Supplementary Information}
% resets enumeration
\setcounter{figure}{0}
\renewcommand{\thefigure}{S\arabic{figure}}
\setcounter{table}{0}
\renewcommand{\thetable}{S\arabic{table}}

\subsection{Models architecture and parameterization}
\label{SI:arch_and_param}

\begin{figure*}[p]
    \centering
    \includegraphics[width=0.95\linewidth]{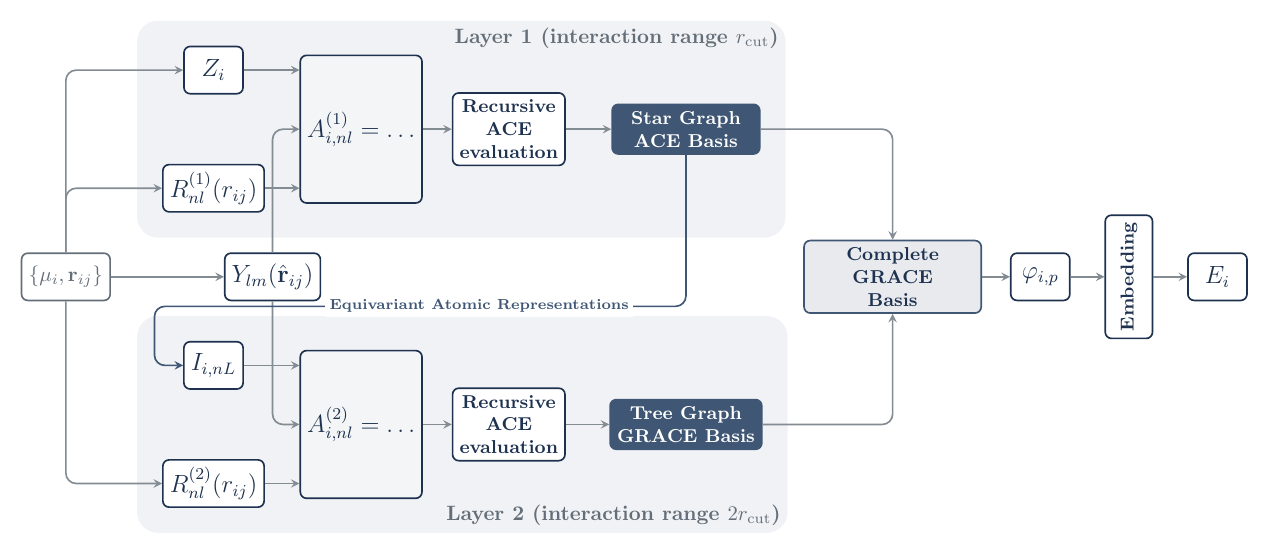}
    \hfill
    \includegraphics[width=0.65\linewidth]{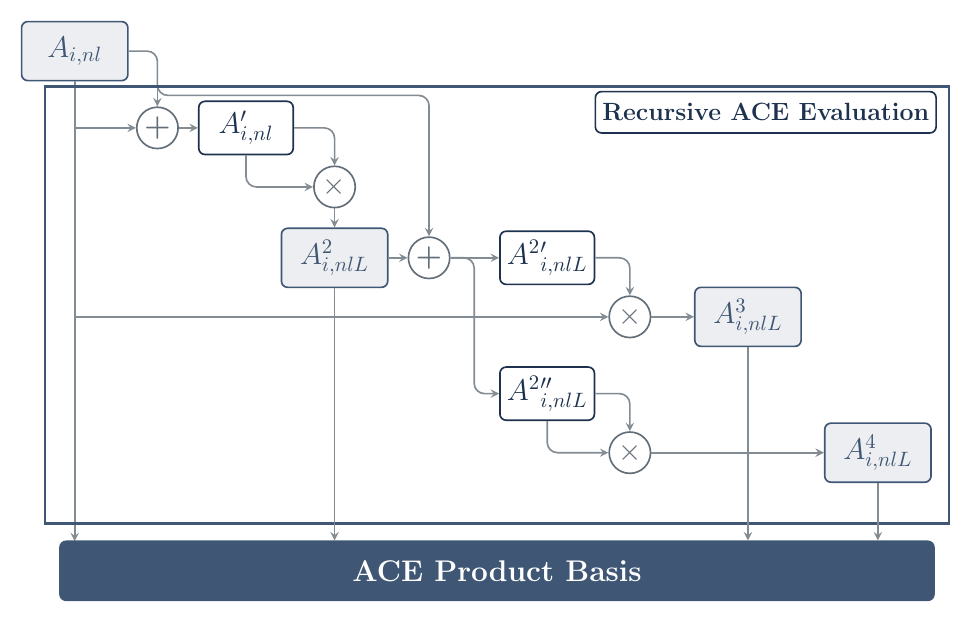}
    \caption{\changed{GRACE models: (top panel) overall scheme and (bottom panel) recursive ACE basis evaluation. See text for more details.}}
    \label{fig:grace-scheme}
\end{figure*}

The configurations of the GRACE foundational potentials \changed{are schematically depicted in Fig.~\ref{fig:grace-scheme} and provided in more detail} in Table~\ref{tab:grace_configs}.

\changed{The atomic structure is represented by chemical species types $\mu_i$ and atomic positions $\mathbf{r}_i$, which are transformed into bond vectors $\mathbf{r}_\mathrm{ij}$ between neighboring atoms within a 6\,\AA\, cutoff radius.
Geometric information is encoded using a Chebyshev radial basis and spherical harmonics $Y_\mathrm{lm}$ up to $l_\mathrm{max}=4$ \cite{Drautz19,Lysogorskiy21,Bochkarev22}. 
The Chebyshev polynomials are transformed into radial functions $R_\mathrm{nl}^{(\ldots)}$ using a multi-layer perceptron (MLP) with two hidden layers, each containing 64 units.
Chemical species are embedded into a 128-dimensional space $Z_i$.
Single-particle basis functions are constructed from these inputs and summed into atomic bases $A_\mathrm{i,nl}$ \cite{Drautz19,bochkarev2024}.
Many-body basis functions up to the fourth product order are constructed via recursive ACE evaluation using sparse coupling operations (see Fig.~\ref{fig:grace-scheme}, bottom panel).
During the recursive ACE basis evaluation, we utilize an equivariant sum operation $\oplus$, summing equivariant quantities with the same $l$-character, and a sparse equivariant coupling $\otimes$ operation which employs Clebsch-Gordan coefficients.
We employ a specific coupling order to avoid degenerate product functions.
The maximum angular momentum, $L_\mathrm{max}$, varies depending on the product order.
In the one-layer model, this procedure yields a star-graph ACE basis.
For two-layer models, equivariant basis functions are linearly combined to define equivariant atomic representations $I_\mathrm{i,nL}$~\cite{bochkarev2022multilayer}. 
These representations carry geometric information about the atomic environment, effectively extending the interaction range to $2r_\mathrm{cut}$. They serve as input for a second recursive ACE evaluation, resulting in a tree graph GRACE basis~\cite{bochkarev2024}.}
Finally, all invariant basis functions from all product orders of the first and second layers \changed{are aggregated to form the complete GRACE basis, which is then} linearly combined to generate a set of atomic densities \changed{$\varphi_{i,p}$}.
The atomic energy is computed \changed{via an embedding scheme} as the sum of the first atomic density and the result of processing the remaining densities through another MLP.
The total energy of the structure is the sum of all atomic energies.
Forces are computed via the gradient of the total energy with respect to the atomic positions.
All calculations are performed in double precision (FP64).
The models are just-in-time (JIT) compiled using XLA to achieve optimal performance.
Further details on the GRACE models will be published in a separate work.

\begin{table*}[t!]
    \centering
    \caption{Configurations of GRACE foundational potentials. See text for more details.}
    \label{tab:grace_configs}
    \begin{tabular}{|c|ccc|ccc|}
        \hline
        Configuration & 1L & 1L-medium & 1L-large & 2L & 2L-medium & 2L-large \\
        \hline
        r$_\mathrm{cut}$ (\AA) & 6 & 6 & 6 & 6 & 6 & 6 \\
        Radial basis func.& Cheb & Cheb & Cheb & Cheb & Cheb & Cheb \\
        Num. radial basis & 8 & 10 & 10 & 8 & 10 & 10 \\
        l$_\mathrm{max}$ & 4 & 4 & 4 & 4 & 4 & 4 \\
        Num.elements & 89 & 89 & 89 & 89 & 89 & 89 \\
        Chem. embedding & 128 & 128 & 128 & 128 & 128 & 128 \\
        Num.radial funcs. & 32 & 32 & 32 & 32 & 42 & 42 \\
        Product order & 4 & 4 & 4 & 4 & 4 & 4 \\
        L$_\mathrm{max}$ per product order (layer 1):  & 4/4/0/0 & 4/4/0/0 & 4/4/0/0 & 4/4/1/0 & 4/4/1/1 & 4/4/3/1 \\
        Tot. num. equivar. funcs. & - &- &- & 4000 & 9576 & 13860 \\
        L$_\mathrm{max}$ per product order (layer 2) & - & - & - & 4/4/0/0 & 3/3/0/0 & 3/3/0/0  \\
        Tot. num. invar. funcs. & 2848 & 2848 & 5664 & 5696 & 7194 & 7194 \\
        Num. densities & 12+1 & 16+1 & 16+1 & 12+1 & 16+1 & 16+1 \\
        Tot. num. params. & 3447148 & 4461497 & 8953529 & 12597516 & 21764956 & 26394284 \\
        \hline
    \end{tabular}
\end{table*}

The GRACE models were parameterized using the recently published OMat24~\cite{barroso_omat24} dataset in combination with the sAlex~\cite{schmidt2023machine,wang2023symmetry} dataset and the MPTraj dataset (v2022.10.28)~\cite{jain2013commentary,deng2023chgnet}.
OMat24 includes a wide range of structures, whereas sAlex and MPTraj include only relaxation trajectories. 
Raw VASP energies, forces and stresses were used for parameterizations.

We employed a loss function that consists of different parts,
\begin{equation}
\Loss = \alpha_\mathrm{E} \Loss_\mathrm{E} + \alpha_\mathrm{F} \Loss_\mathrm{F} + \alpha_\mathrm{S} \Loss_\mathrm{S},
\end{equation}
where $\Loss_\mathrm{E}$, $\Loss_\mathrm{F}$, and $\Loss_\mathrm{S}$ correspond to losses of energy per atom, force component, and stress component, respectively.
We utilize the Huber loss for $\Loss$ with parameter $\delta=0.01$ for all components.
All models were initially trained on the OMat24 dataset for 12 epochs for 1L models and for 8 epochs for 2L models, constituting the "OMat-base" models. For this stage we use $\alpha_\mathrm{E}:\alpha_\mathrm{F}:\alpha_\mathrm{S} = 16:128:128$. Subsequently, we fine-tuned "OMat-base" models on the combination of MPTraj and sAlex datasets for additional 8 and 4 epochs for 1L and 2L models respectively with $\alpha_\mathrm{E}:\alpha_\mathrm{F}:\alpha_\mathrm{S} = 128:128:256$, leading to the "OAM" models. In addition, we fine-tuned "OMat-base" models on the same OMat24 dataset for additional 2 epochs with adjusted loss component weights to $\alpha_\mathrm{E}:\alpha_\mathrm{F}:\alpha_\mathrm{S} = 128:128:256$ to give more weight to energies. This models are designated with "ft-E".

For loss optimization we employed the Adam~\cite{kingma2014adam} optimizer with cosine learning rate reduction scheme, initial learning rate of 8$\times10^{-3}$ and minimum learning rate of 5$\times10^{-4}$.
For fine tuning we use constant learning rate of 1$\times10^{-4}$.
To optimize data throughput we split the data into batches based on the total number of bonds rather than structures. The batch size was set to 165000 bonds per device which on average corresponds to about 200 structures.
Training was performed using a single node with 8 Nvidia H100 80GB GPUs. 
Complete training cost of GRACE models varies between 400 and 700 GPU hours for the smallest and the most complex model, respectively.

\subsection{\changed{Model Distillation}}

\begin{figure*}[t!]
\centering
\includegraphics[width=\linewidth]{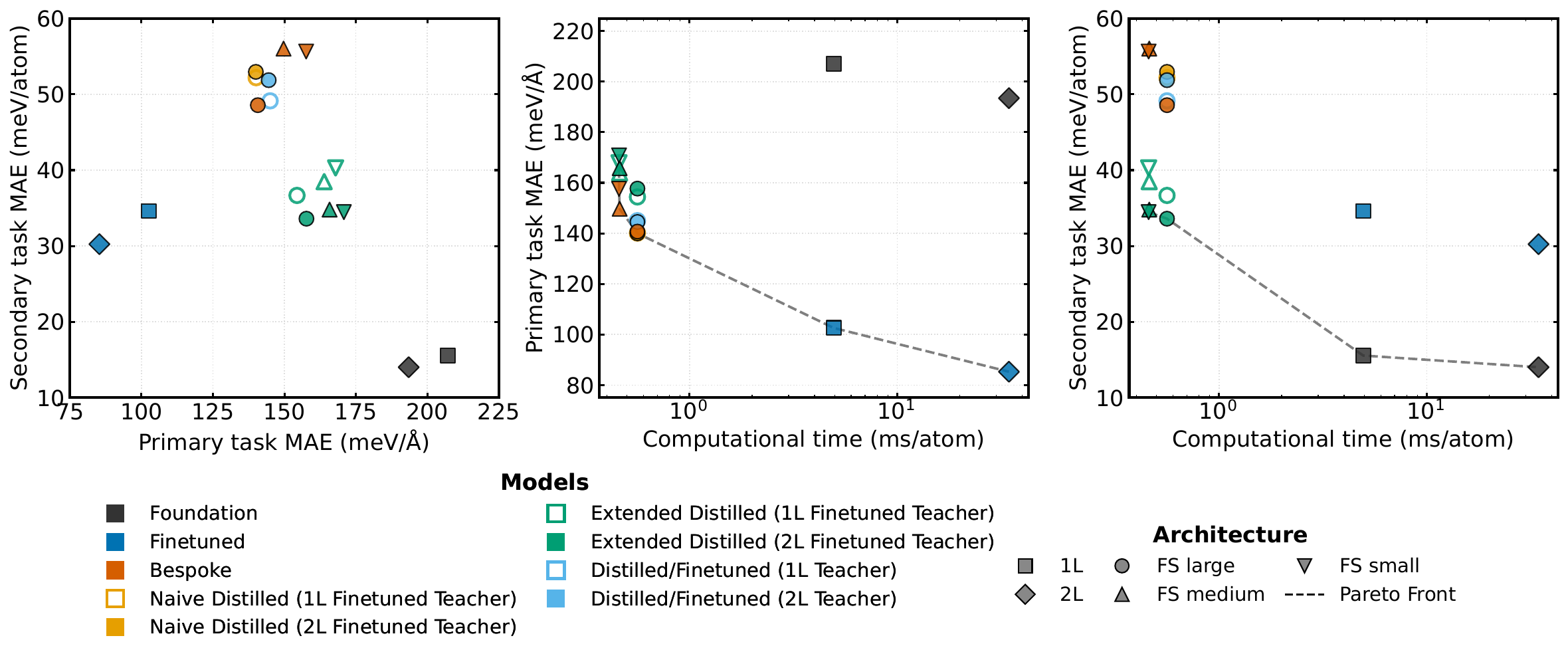}
\caption{
\changed{
Performance trade-offs for GRACE models trained via different strategies. 
{Left:} Accuracy on the  primary task (MAE of forces for HEA25S validation set) versus accuracy on the secondary task (MAE of formation energy of unaries and binaries from Materials Project).
Lower values on both axes indicate better performance.
{Middle and Right:} Computational cost (wall-clock time per atom step) measured on a single Intel(R) Xeon(R) W-2255 CPU  with 10 physical cores, as normalized per core versus primary  (middle) and secondary  (right) tasks accuracies.
The Pareto front (dashed line) highlights the optimal trade-off between speed and accuracy. Symbols differentiate model architectures (squares/diamond: GRACE-1L/2L models;  and GRACE-FS models: circles - large, up triangles - medium, down triangles - small), while colors indicate the training methodology. Empty symbols indicate distilled models with 1L teacher, whereas filled symbols - 2L teacher.
}
}
\label{fig:distill_three}
\end{figure*}

\changed{

Multiple fine-tuning and distillation pathways were evaluated, as detailed in Table~\ref{tab:distill} and illustrated in Figure~\ref{fig:distill_three}. 

Computational performance benchmarks were conducted on an Intel(R) Xeon(R) W-2255 CPU (3.70\,GHz) equipped with 10 physical cores. We measured the wall-clock run time within LAMMPS for a system containing 2000 tungsten atoms in a body-centered cubic (BCC) configuration.
For GRACE-2L models, LAMMPS MPI parallelization was disabled; instead, we relied on TensorFlow's internal parallelization with XLA compilation to utilize the available cores. The environment variables were configured as follows: \texttt{OMP\_NUM\_THREADS=1}, \texttt{TF\_NUM\_INTRAOP\_THREADS=20}, and \texttt{TF\_NUM\_INTEROP\_THREADS=1}.
For GRACE-1L models, a hybrid strategy was employed to saturate CPU capacity, combining LAMMPS MPI parallelization (10 MPI ranks) with XLA/TensorFlow internal parallelization, utilizing \texttt{TF\_NUM\_INTRAOP\_THREADS=8} and \texttt{TF\_NUM\_INTEROP\_THREADS=1}. Finally, for GRACE-FS, parallelization was achieved exclusively using 10 MPI ranks in LAMMPS.
For comparison, we also measured computational performance on a single NVIDIA A100 80Gb GPU.
GRACE-FS models of three different complexities were considered: large, medium, and small. 
We initiated our study with both GRACE-2L-OMAT and GRACE-1L-OMAT foundation models.
These models were fine-tuned on the HEA25S dataset, resulting in the 2L-OMAT-HEA25S-ft and 1L-OMAT-HEA25S-ft models. As expected, the 2L model demonstrated better metrics after fine-tuning compared to the 1L model.
These models were then distilled into two corresponding GRACE-FS-large models (FS-large-distill-1L/2L-OMAT-ft-HEA25s), implementing a naive distillation scenario. Interestingly, the accuracy of the distilled models does not depend on the distillation source, suggesting that the student models have reached their learning capacities.
We also parameterized three bespoke GRACE-FS models of varying complexities. As expected, the accuracy of these models increases with model complexity.
Additionally, we evaluated another sequence: first distilling the foundation models into student GRACE-FS models using the HEA25S dataset, followed by fine-tuning the students using the original HEA25S dataset with DFT references. As shown in the results, all three pathways yielded models of similar accuracy in both primary and secondary tasks, although the latter performed slightly worse compared to the foundation and fine-tuned models.
To improve the performance of distilled models on the secondary task, we extended the distillation dataset with unary and binary compound structures (both ideal and rattled). This allowed us to parameterize GRACE-FS models across three different complexities, reaching secondary task accuracy levels comparable to fine-tuned models while only slightly compromising accuracy on the primary task. Increasing the complexity of both student and teacher models systematically improved metrics on both primary and secondary tasks (Fig.~\ref{fig:distill_three}, left panel).
The Pareto front of speed-accuracy, as shown in Fig.~\ref{fig:distill_three} (middle and right panels), is formed by  the fine-tuned foundation 1L- and 2L-models, as well as the distilled models.

Finally, we performed an extended distillation where, in addition to total energies, forces, and stresses, atomic site energies were included as reference targets (FS-large-extdistill-atomic-energy-2L-OMAT-ft-HEA25s). However, the resulting metrics were slightly worse than those obtained from the identical parameterization without atomic energies (FS-large-extdistill-2L-OMAT-ft-HEA25s).
}

\begin{table*}[ht]
\centering
\renewcommand{\arraystretch}{1.1}
\caption{\changed{Distillation of GRACE models. Error metrics for:
(1) Primary task: Energy (meV/at) and forces (meV/\AA) for HEA25S (PBEsol, non-magnetic), 25 elements.
(2) Secondary task: Formation energies (meV/atom) of unary and binary structures for Materials Project (PBE) composed of 19 non-magnetic elements.
(3) Computational performance $t_\mathrm{LAMMPS}$ ($\mu$s/atom/step/device), as measured per-device (NVIDIA A100 80Gb GPU or Intel(R) Xeon(R) W-2255 CPU @ 3.70GHz with 10 physical cores).
See text for details.
}}
\setlength{\tabcolsep}{3pt}
\resizebox{\textwidth}{!}{%
  \begin{tabular}{|l|c|c|c|c|c|c|c|c|}
  \hline
  \multirow{2}{*}{Model} & \multirow{2}{*}{Training set} & \multirow{2}{*}{Reference method} & \multicolumn{2}{c|}{HEA25S} & \multicolumn{2}{c|}{MatProj} & \multicolumn{2}{c|}{\changed{$t_\mathrm{LAMMPS}$}} \\
  \cline{4-5} \cline{6-7} \cline{8-9}
  & & & E-MAE & {F-MAE} & unary & {binary} & A100 & 10xCPU \\
  \hline
  \multicolumn{9}{|c|}{\textbf{Foundation}} \\
  \hline
  2L-OMAT & OMat24 & PBE & 617.9 & 193.5 &\textbf{ 19.7} & \textbf{14} & 25 & 3448 \\
  1L-OMAT & OMat24 & PBE & 615 & 207.1 & 23.9 & 15.6 & 10 & 494 \\
  \hline
  \hline
  \multicolumn{9}{|c|}{\textbf{Finetuned}} \\
  \hline
  2L-OMAT-HEA25S-ft & HEA25S & PBEsol & \textbf{3.5} & \textbf{85} & 23 & 30 & 25 & 3448 \\
  1L-OMAT-HEA25S-ft & HEA25S & PBEsol & 4.1 & 102.6 & 28.0 & 34.6 & 10 & 494 \\
  \hline
  \hline
  \multicolumn{9}{|c|}{\textbf{Naive distilled}} \\
  \hline
  FS-large-distill-2L-OMAT-ft-HEA25s & HEA25S & 2L-OMAT-ft-HEA25s & \textbf{7.5} & \textbf{140.0} & 34.1 & 53.0 & 8 & 61 \\
  FS-large-distill-1L-OMAT-ft-HEA25s & HEA25S & 1L-OMAT-ft-HEA25s & \textbf{7.5} & 140.1 & 34.1 & 52.2 & 8 & 61 \\
  \hline
  \multicolumn{9}{|c|}{\textbf{Bespoke}} \\
  \hline
  FS-large-HEA25s & HEA25S & PBEsol & 7.6 & 140.7 & 30.71 & 48.56 & 8 & 61 \\
  FS-medium-HEA25s & HEA25S & PBEsol & 9.4 & 149.7 & 34.9 & 56.0 & 8 & 46 \\
  FS-small-HEA25s & HEA25S & PBEsol & 9.9 & 157.6 & 35.7 & 55.6 & 8 & 46 \\
  \hline
  \multicolumn{9}{|c|}{\textbf{Distilled/Finetuned}} \\
  \hline
  FS-large-ft-HEA-2L-OMAT & HEA25S & 2L-OMAT & - & 219.9 & 32.8 & 43.8 & 8 & 61 \\
  FS-large-ft-HEA-2L-OMAT-ft-HEA25s & HEA25S & PBEsol & 8.2 & 144.5 & 34.1 & 51.9 & 8 & 61 \\
  FS-large-ft-HEA-1L-OMAT & HEA25S & 1L-OMAT & - & 224.2 & 31.4 & 39.9 & 8 & 61 \\
  FS-large-ft-HEA-1L-OMAT-ft-HEA25s & HEA25S & PBEsol & 8.3 & 145.0 & 33.1 & 49.1 & 8 & 61 \\
  \hline
  \multicolumn{9}{|c|}{\textbf{Extended distilled}} \\
  \hline
  {FS-large-extdistill-2L-OMAT-ft-HEA25s} & ext. HEA25S & 2L-OMAT-ft-HEA25s & 9.9 & 158 & \textbf{28.5} & \textbf{33.6} & 8 & 61 \\
  FS-large-extdistill-1L-OMAT-ft-HEA25s & ext. HEA25S & 1L-OMAT-ft-HEA25s & 10.1 & 154.4 & 30.3 & 36.7 & 8 & 61 \\
  FS-medium-extdistill-2L-OMAT-ft-HEA25s & ext. HEA25S & 2L-OMAT-ft-HEA25s & 11.5 & 165.8 & 30.7 & 34.8 & 8 & 46 \\
  FS-medium-extdistill-1L-OMAT-ft-HEA25s & ext. HEA25S & 1L-OMAT-ft-HEA25s & 11.6 & 163.9 & 30.2 & 38.5 & 8 & 46 \\
  FS-small-extdistill-2L-OMAT-ft-HEA25s & ext. HEA25S & 2L-OMAT-ft-HEA25s & 12.0 & 170.8 & 30.8 & 34.5 & 8 & 46 \\
  FS-small-extdistill-1L-OMAT-ft-HEA25s & ext. HEA25S & 1L-OMAT-ft-HEA25s & 11.6 & 167.9 & 32.2 & 40.3 & 8 & 46 \\
  \hline
  \multicolumn{9}{|c|}{\textbf{Extended distilled (atomic energies)}} \\
  \hline
  {FS-large-extdistill-atomic-energy-2L-OMAT-ft-HEA25s} & ext. HEA25S (+atomic energies) & 2L-OMAT-ft-HEA25s & 10.9 & 159.6 & 29.3 & 34.1 & 8 & 61 \\
  \hline
  \end{tabular}%
}
\label{tab:distill}
\end{table*}

% \clearpage
\subsection{Molecular dynamics stability and energy conservation}

\begin{figure*}[h!]
\centering
\includegraphics[width=0.75\linewidth]{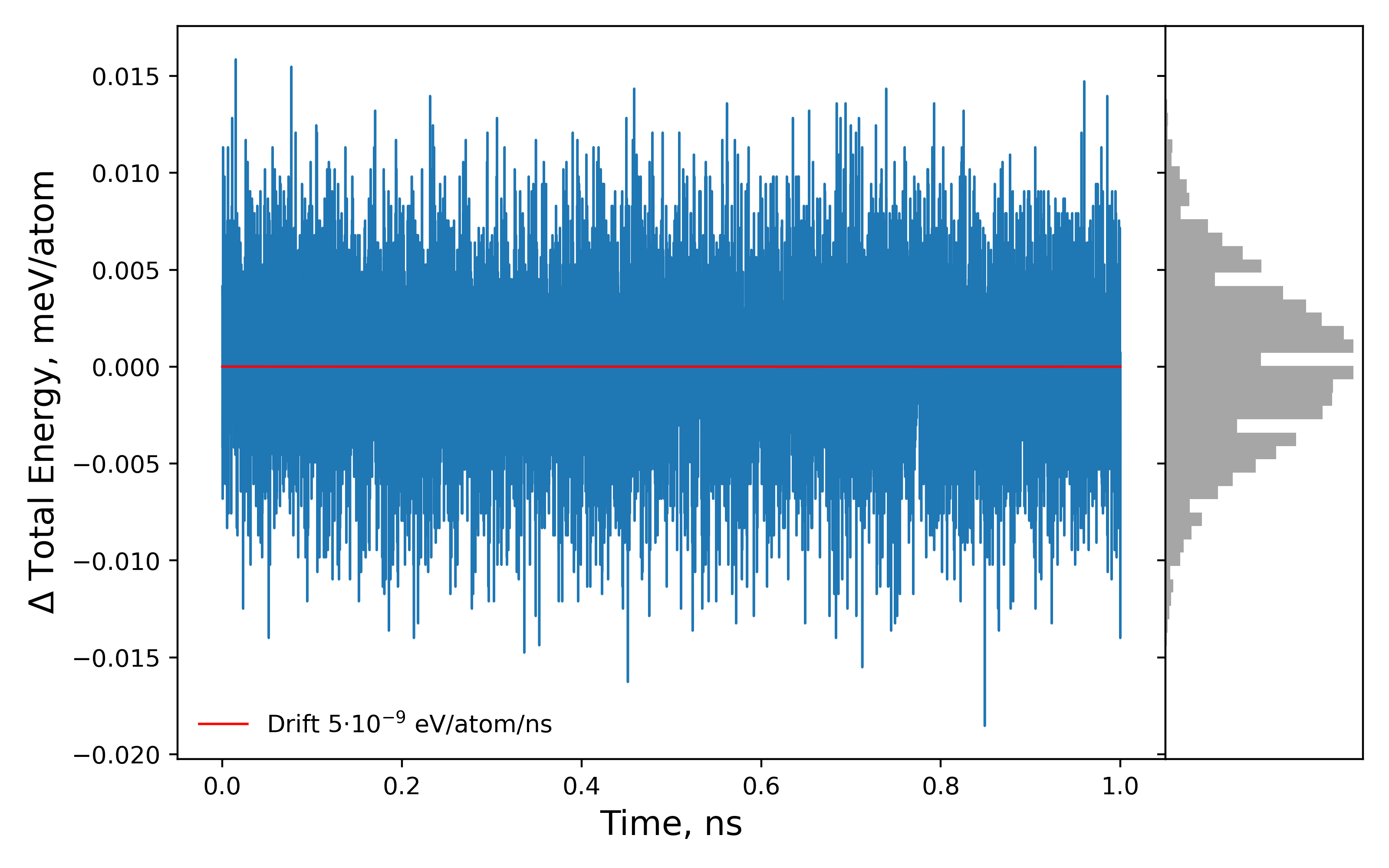}
\caption{
\protect\changed{
Total energy conservation for a 3,000-atom FLiBe supercell simulated in the NVE ensemble at 973\,K using the GRACE-2L-OMAT-L model. The main plot shows the deviation of the total energy per atom from the mean over a 1\,ns trajectory. The red line indicates the linear regression of the energy drift, calculated as $5\cdot10^{-9}$~eV/atom/ns. The side panel shows the corresponding histogram of energy fluctuations, demonstrating a stable equilibrium distribution.}
}
\label{fig:drift}
\end{figure*}

\changed{To assess the long-term numerical stability of the foundational models, we analyzed the conservation of total energy during a 1\,ns molecular dynamics simulation of molten FLiBe in the NVE ensemble at 973\,K. 
Figure illustrates the time evolution of the total energy fluctuations per atom. 
The system exhibits exceptional stability, characterized by a negligible linear drift of $5\cdot10^{-9}$~eV/atom/ns. 
Furthermore, the histogram of energy fluctuations (right panel) displays a symmetric, Gaussian - like distribution centered at zero, confirming the absence of systematic  heating or cooling artifacts over the simulation trajectory.}

\clearpage
\subsection{Detailed Validation Benchmarks}
This section provides a detailed overview of the validation benchmarks. 

\begin{figure*}[t!]
\centering
\includegraphics[width=0.5\linewidth]{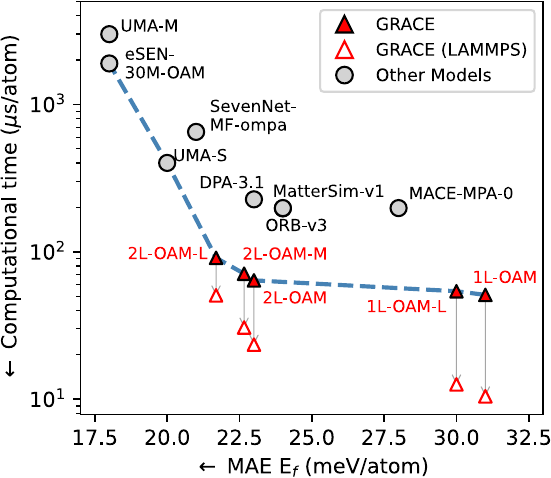}
\caption{
Model performance for MAE of formation energy (from MatBench Discovery benchmark) versus computational time per atom.
The blue dashed line links Pareto optimal models. Computational performance is estimated via ASE (filled symbols) and LAMMPS (open symbols), with GRACE models indicated in red.
}
\label{fig:two_pareto_mae}
\end{figure*}

\begin{figure*}[t!]
\centering
\includegraphics[width=0.5\linewidth]{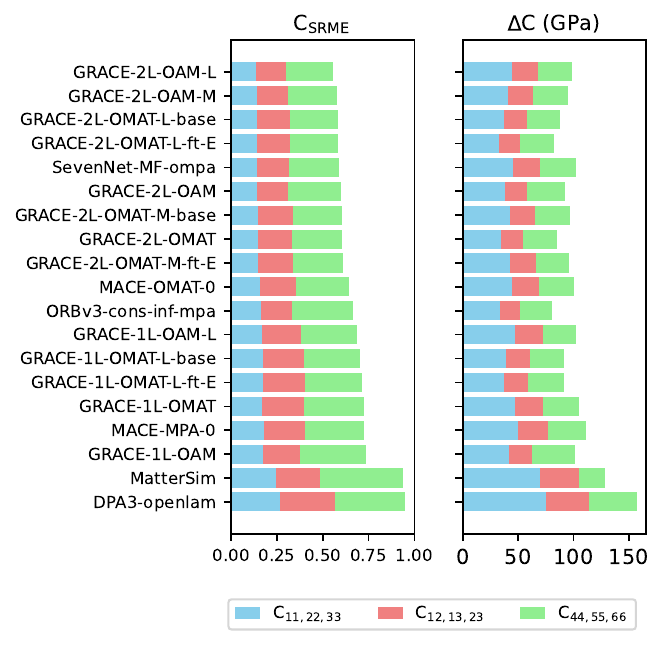}
\caption{SRME and MAE for three different subgroups of elastic tensor, referenced to Materials Project~\cite{de2015charting}}
\label{fig:elast_long}
\end{figure*}

\begin{figure*}[t]
\centering
\includegraphics[width=0.5\linewidth]{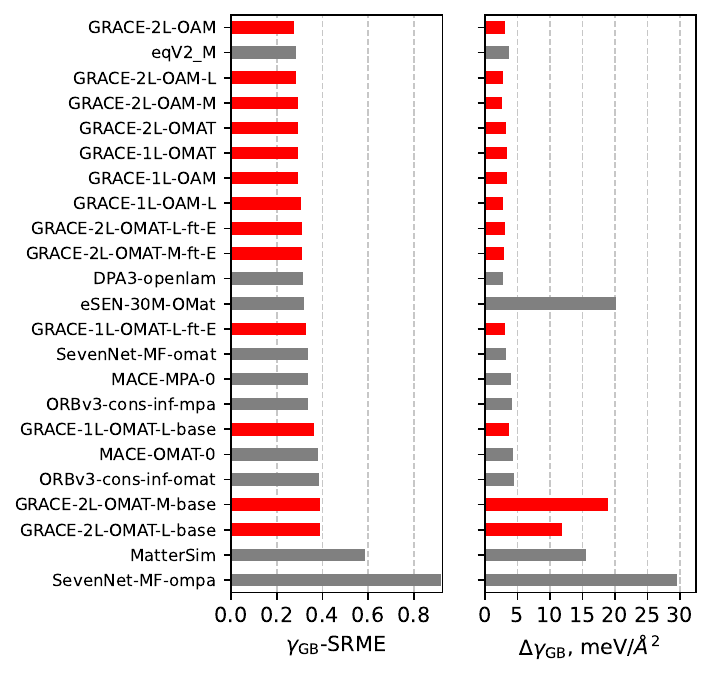}
\caption{Error metrics for unary grain boundaries formation energies: $\gamma_\mathrm{GB}$-SRME (left) and mean absolute error $\Delta \gamma_\mathrm{GB}$ (right). GRACE models are highlighted in red.
}
\label{fig:GBs_long}
\end{figure*}

\begin{figure*}[t]
\centering
\includegraphics[width=\linewidth]{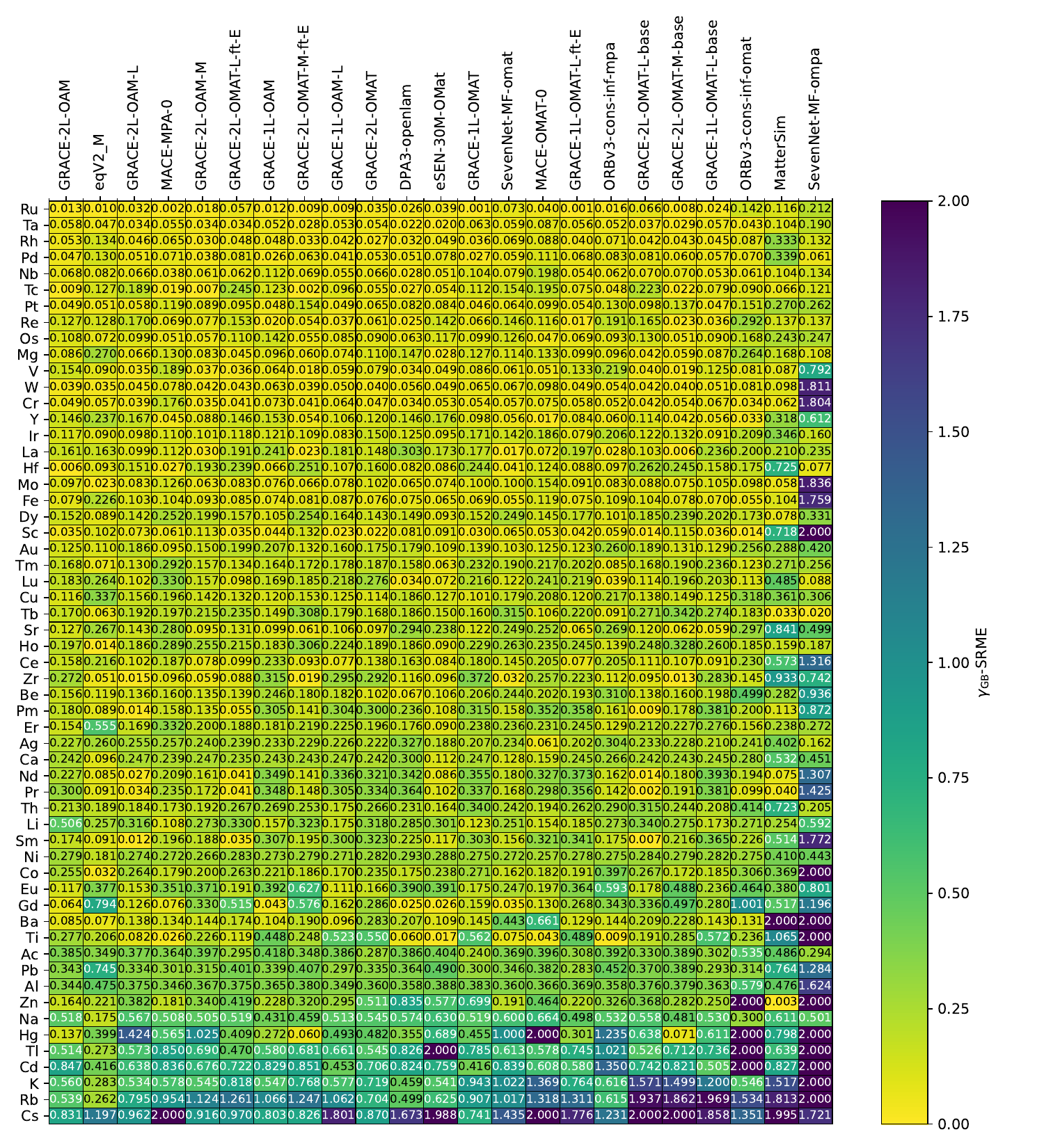}
\caption{$\gamma_\mathrm{GB}$-SRME decomposed into different elements and MLIPs.}
\label{fig:GB_table}
\end{figure*}

\begin{figure*}[t]
\centering
\includegraphics[width=0.5\textwidth]{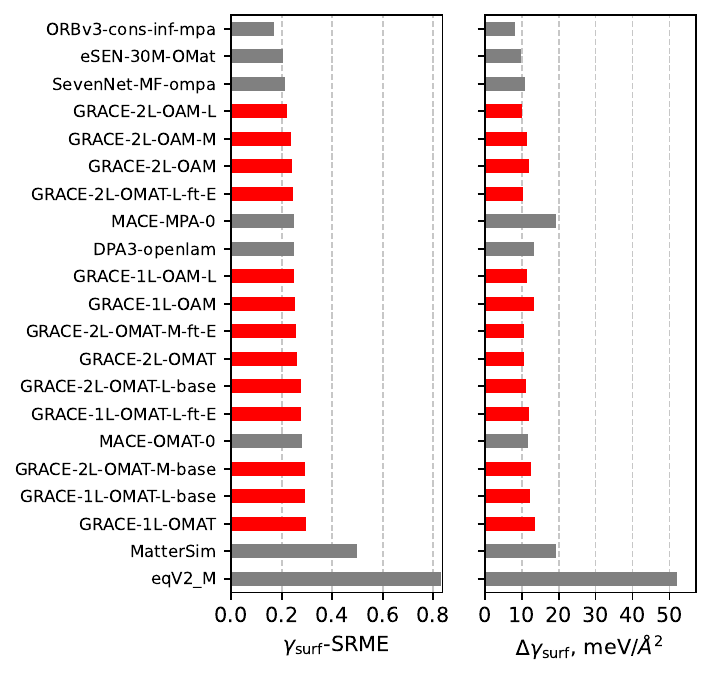}
\caption{Error metrics for unary surface formation energies: $\gamma_\mathrm{surf}$-SRME (left) and mean absolute error $\Delta \gamma_\mathrm{surf}$ (right). GRACE models are highlighted in red.
}
\label{fig:surfs_long}
\end{figure*}

\begin{figure*}[t]
\centering
\includegraphics[width=1\linewidth]{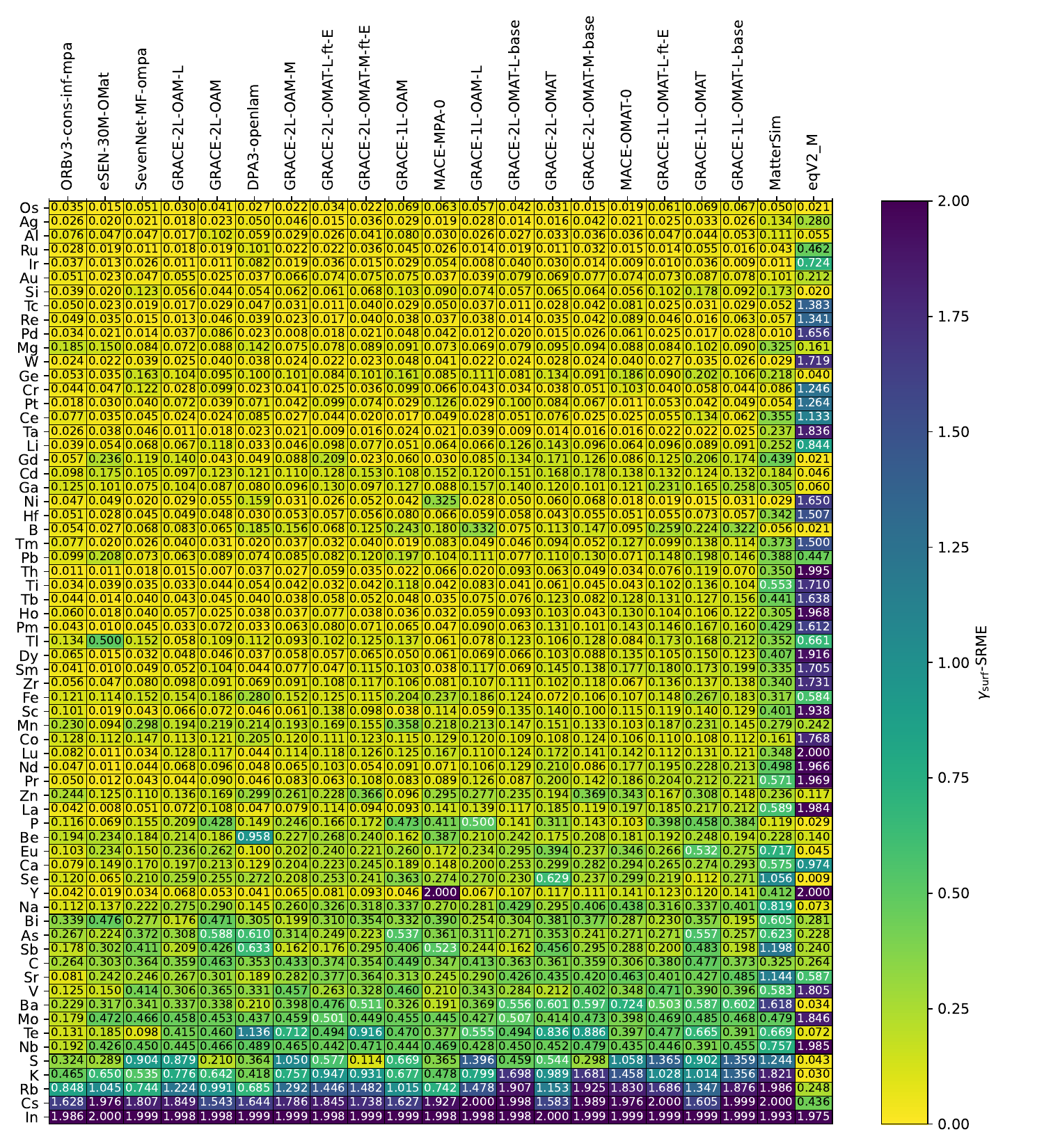}
\caption{$\gamma_\mathrm{surf}$-SRME decomposed into different elements and MLIPs.}
\label{fig:SURF_table}
\end{figure*}

\begin{figure*}[t]
\centering

\includegraphics[width=0.5\textwidth]{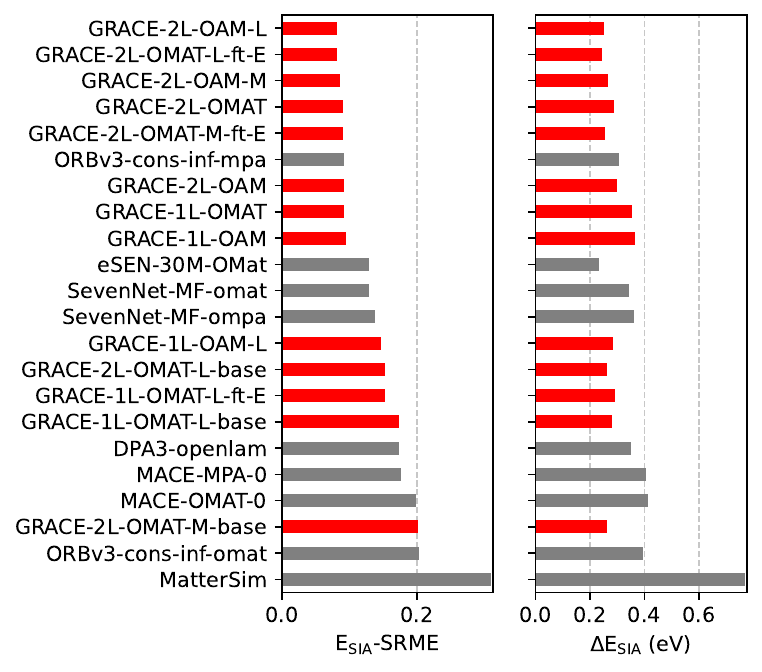}
\caption{Error metrics for SRME of point defects formation energies in unaries: self-interstitials and vacancies. GRACE models are highlighted in red.
}
\label{fig:SIA_delta_long}
\end{figure*}

\begin{figure*}[t]
\centering
\includegraphics[width=0.5\textwidth]{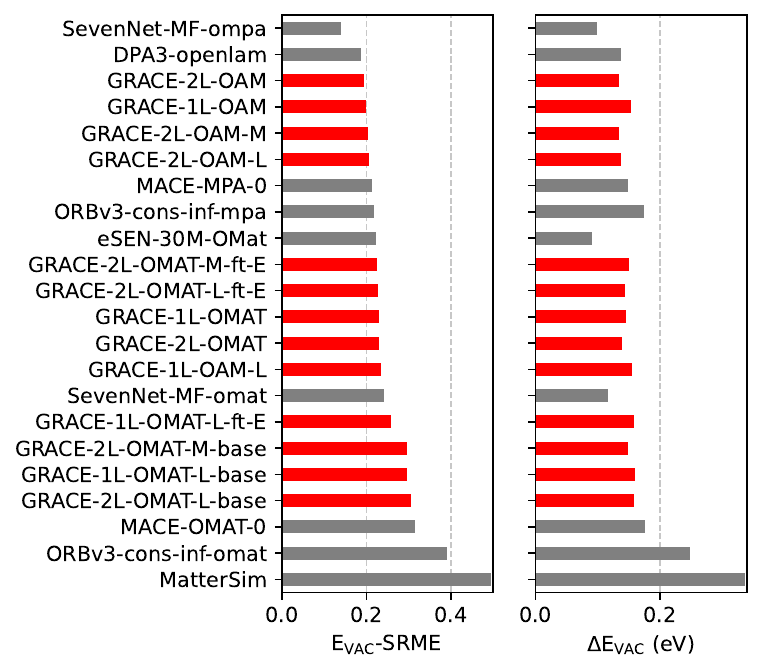}
\caption{Error metrics for SRME of point defects formation energies in unaries: self-interstitials and vacancies. GRACE models are highlighted in red.
}
\label{fig:VAC_delta_long}
\end{figure*}

\begin{figure*}[t]
\centering

\includegraphics[height=0.9\textheight, keepaspectratio]{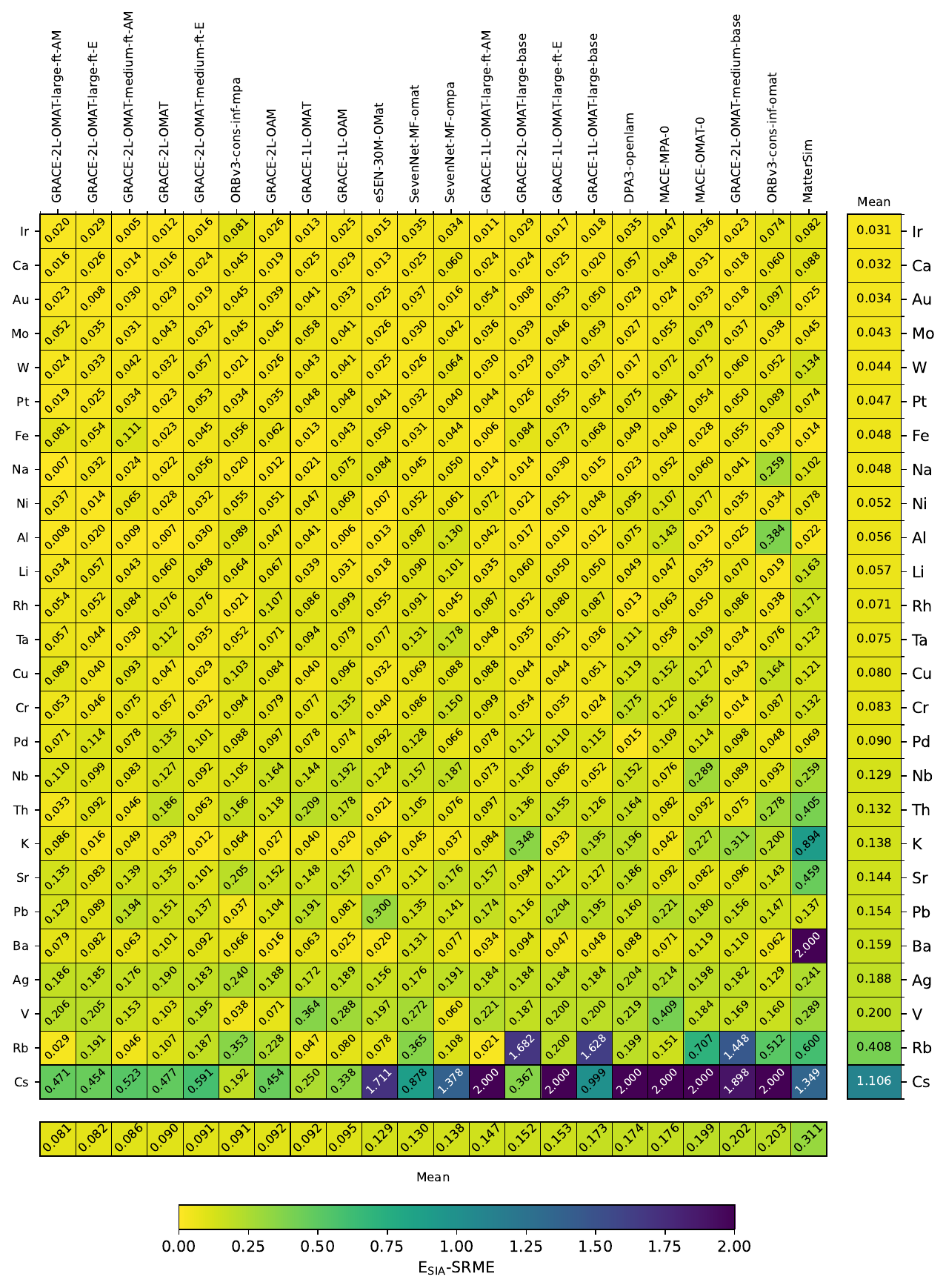}
\caption{E$_\mathrm{SIA}$-SRME  decomposed into elements and MLIPs}
\label{fig:SIA_srme_long}
\end{figure*}

\begin{figure*}[t]
\centering

\includegraphics[height=0.9\textheight, keepaspectratio]{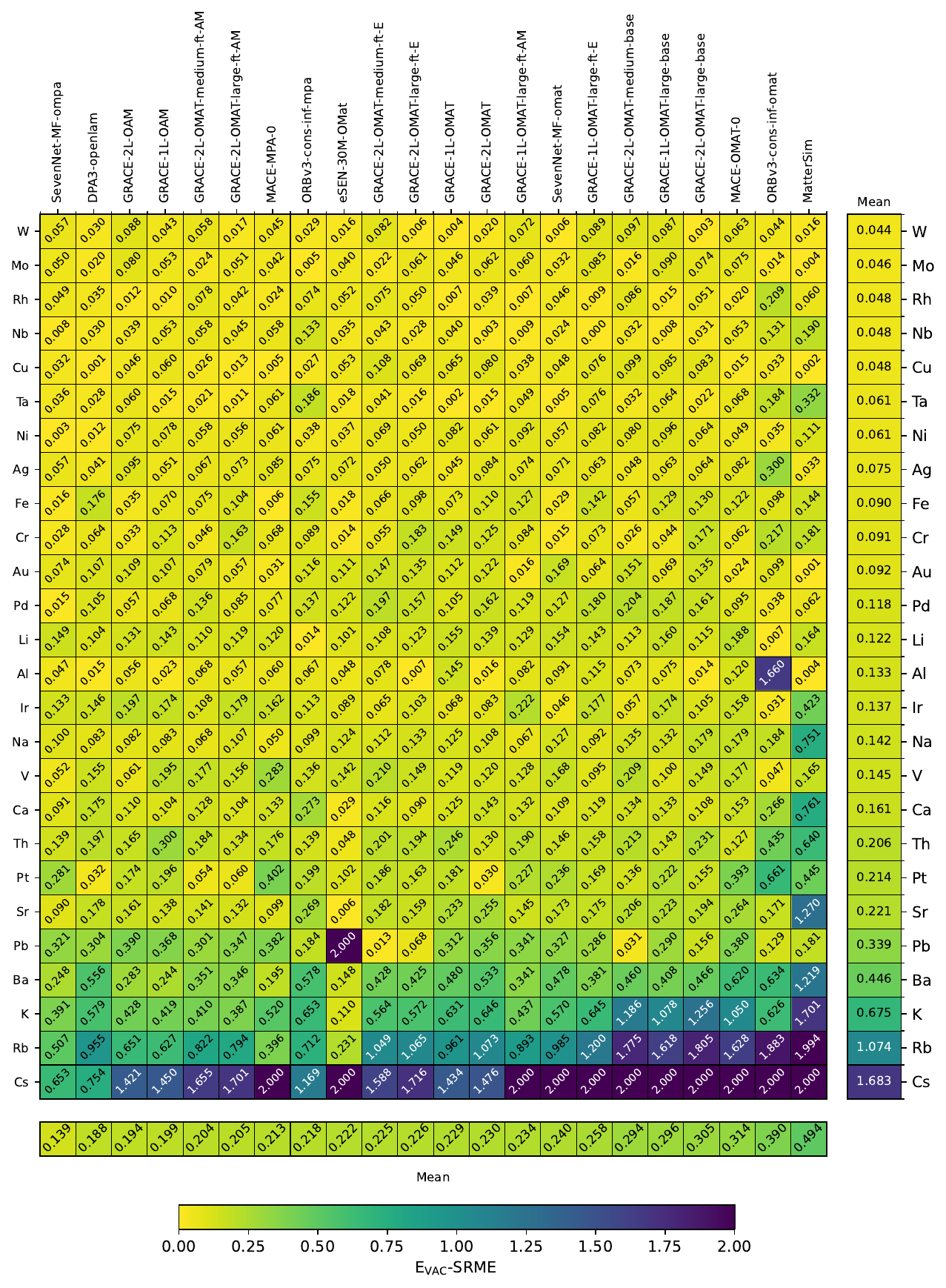}
\caption{E$_\mathrm{vac}$-SRME  decomposed into elements and MLIPs}
\label{fig:VAC_srme_long}
\end{figure*}

Mean absolute error of formation energy from Matbench Discovery vs. computational performance are shown on Fig.~\ref{fig:two_pareto_mae}.
The performance for predicting the elastic tensor is shown in Fig.~\ref{fig:elast_long}. For structural defects, error metrics for unary grain boundary formation energies are presented in Fig.~\ref{fig:GBs_long}, with a detailed element-wise decomposition of the SRME provided in Fig.~\ref{fig:GB_table}. Similarly, the performance for unary surface formation energies is summarized in Fig.~\ref{fig:surfs_long}, with a corresponding elemental breakdown shown in Fig.~\ref{fig:SURF_table}. Finally, the model's accuracy for point defects in unaries is presented for self-interstitials (Fig.~\ref{fig:SIA_delta_long}) and vacancies (Fig.~\ref{fig:VAC_delta_long}). The element-decomposed SRME for these point defects are detailed in Fig.~\ref{fig:SIA_srme_long} and Fig.~\ref{fig:VAC_srme_long}, respectively.

\end{document}